\documentclass[useAMS,usenatbib]{mn2e}
\usepackage{amsmath}
\usepackage{graphics}
\usepackage{graphicx}
\usepackage{times}
\usepackage{times}
\usepackage{tabularx}

\newcommand{\cpp}{c_{\perp}}
\newcommand{\cll}{c_{||}}
\newcommand{\dpp}{d_{\perp}}
\newcommand{\gmod}{g_{\rm mod}}
\newcommand{\rmod}{r_{\rm mod}}
\newcommand{\imod}{i_{\rm mod}}

\newcommand{\rcmod}{r_{\rm cmod}}
\newcommand{\icmod}{i_{\rm cmod}}
\newcommand{\ipsf}{i_{\rm psf}}
\newcommand{\zpsf}{z_{\rm psf}}
\newcommand{\zmod}{z_{\rm mod}}
\newcommand{\rpsf}{r_{\rm psf}}
\newcommand{\ifib}{i_{\rm fib2}}

\def\simlt{\stackrel{<}{{}_\sim}}

\begin{document}

\title[BOSS LSS targeting and catalogues]{SDSS-III Baryon Oscillation Spectroscopic Survey
  Data Release 12: galaxy target selection and large scale structure catalogues}
\author[Reid et al.]{\parbox{\textwidth}{
Beth Reid$^{1,2}$, 
Shirley Ho$^{3}$, 
Nikhil Padmanabhan$^{4}$, 
Will J. Percival\thanks{E-mail: will.percival@port.ac.uk}$^{5}$, 
Jeremy Tinker$^{6}$, 
Rita Tojeiro$^{7}$, 
Martin White$^{8,2}$, 
Daniel J. Eisenstein$^{9}$,
Claudia Maraston$^{5}$,
Ashley J. Ross$^{10}$,
Ariel G. S\'anchez$^{11}$,
David Schlegel$^{1,2}$,
Erin Sheldon$^{12}$,
Michael A. Strauss$^{13}$
Daniel Thomas$^{5}$,
David Wake$^{14,15}$,
Florian Beutler$^{2}$,
Dmitry Bizyaev$^{16,17}$,
Adam S. Bolton$^{18}$,
Joel R. Brownstein$^{18}$,
Chia-Hsun Chuang$^{19}$,
Kyle Dawson$^{18}$,
Paul Harding$^{20}$,
Francisco-Shu Kitaura$^{21}$,
Alexie Leauthaud$^{22}$,
Karen Masters $^{5}$,
Cameron K. McBride$^{9}$,
Surhud More$^{22}$,
Matthew D. Olmstead$^{23,18}$,
Daniel Oravetz$^{16}$,
Sebasti\'an~E.~Nuza$^{24}$
Kaike Pan$^{16}$,
John Parejko$^{4}$,
Janine Pforr$^{5,25}$,
Francisco Prada$^{19,26,27}$,
Sergio Rodr\'iguez-Torres$^{19,26,28}$,
Salvador Salazar-Albornoz$^{11,29}$,
Lado Samushia$^{30,31,5}$
Donald P. Schneider$^{32,33}$,
Claudia G. Sc\'occola$^{34,35,36}$,
Audrey Simmons$^{16}$,
Mariana Vargas-Magana$^{37}$
} \vspace*{4pt} \\
$^{1}$ Berkeley Center for Cosmological Physics, Department of Physics, University of California, Berkeley, CA, 94720, USA \\
$^{2}$ Lawrence Berkeley National Laboratory, 1 Cyclotron Road, Berkeley, CA 94720, USA \\
$^{3}$ Bruce and Astrid McWilliams Center for Cosmology, Department of Physics, Carnegie Mellon University, 5000 Forbes Ave, Pittsburgh, PA 15213, USA\\
$^{4}$ Yale Center for Astronomy and Astrophysics, Yale University, New Haven, CT, 06520, USA \\
$^{5}$ Institute of Cosmology \& Gravitation, Dennis Sciama Building, University of Portsmouth, Portsmouth, PO1 3FX, UK\\
$^{6}$ Center for Cosmology and Particle Physics, Department of Physics, New York University, 4 Washington Place, New York, NY 10003, USA\\
$^{7}$ School of Physics and Astronomy, University of St Andrews, North Haugh, St Andrews KY16 9SS, UK\\
$^{8}$ Departments of Physics and Astronomy, University of California, Berkeley, CA 94720, USA\\
$^{9}$ Harvard-Smithsonian Center for Astrophysics, 60 Garden St., Cambridge, MA 02138, USA\\
$^{10}$ Center for Cosmology and AstroParticle Physics, The Ohio State University, Columbus, OH 43210, USA\\
$^{11}$ Max-Planck-Institut f\"ur extraterrestrische Physik, Postfach 1312, Giessenbachstr., 85741 Garching, Germany\\
$^{12}$ Brookhaven National Laboratory, Bldg 510, Upton, New York, NY 11973, USA\\
$^{13}$ Department of Astrophysical Sciences, Princeton University, Princeton, NJ 08544 USA\\
$^{14}$ Department of Astronomy, University of Winsconsin-Madison, 475 N. Charter Street, Madison, WI 53706-1582, USA\\
$^{15}$ Department of Physical Sciences, The Open University, Milton Keynes MK7 6AA, UK\\
$^{16}$ Apache Point Observatory and New Mexico State University, P.O. Box 59, Sunspot, NM, 88349-0059, USA\\
$^{17}$ Sternberg Astronomical Institute, Moscow State University, Moscow\\
$^{18}$ Department Physics and Astronomy, University of Utah, 115 S 1400 E, Salt Lake City, UT 84112, USA\\
$^{19}$ Instituto de F\'{\i}sica Te\'orica, (UAM/CSIC), Universidad Autonoma de Madrid, Cantoblanco, E-28049 Madrid, Spain\\
$^{20}$ Department of Astronomy, Case Western Reserve University, 10900 Euclid Ave, Cleveland, OH 44106, USA\\
$^{21}$ Leibniz-Institut f\"ur Astrophysik Potsdam (AIP), An der Sternwarte 16, D-14482 Potsdam, Germany\\
$^{22}$ Kavli IPMU (WPI), UTIAS, The University of Tokyo, Kashiwa, Chiba 277-8583, Japan\\
$^{23}$ King's College, Department of Chemistry and Physics, Wilkes Barre, PA 18711, USA\\
$^{24}$ Leibniz-Institut f\"ur Astrophysik Potsdam (AIP), An der Sternwarte 16, D-14482 Potsdam, Germany\\
$^{25}$ Aix Marseille Université, CNRS, LAM (Laboratoire d'Astrophysique de Marseille) UMR 7326, 13388, Marseille, France\\
$^{26}$ Campus of International Excellence UAM+CSIC, Cantoblanco, E-28049 Madrid, Spain\\
$^{27}$ Instituto de Astrofisica de Andaluca (CSIC), Glorieta de la Astronoma, E-18080 Granada, Spain\\
$^{28}$ Departamento de F{\'i}sica Te{\'o}rica, Universidad Aut{\'o}noma de Madrid, Cantoblanco, 28049, Madrid, Spain\\
$^{29}$ Universit\"ats-Sternwarte M\"unchen, Scheinerstrasse 1, 81679 Munich, Germany\\
$^{30}$ Kansas State University, Manhattan KS 66506, USA\\
$^{31}$ National Abastumani Astrophysical Observatory, Ilia State University, 2A Kazbegi Ave., GE-1060 Tbilisi, Georgia\\
$^{32}$ Department of Astronomy and Astrophysics, The Pennsylvania State University, University Park, PA 16802 USA\\
$^{33}$ Institute for Gravitation and the Cosmos, The Pennsylvania State University, University Park, PA 16802 USA\\
$^{34}$ Instituto de Astrof{\'\i}sica de Canarias (IAC), C/V{\'\i}a L\'actea, s/n, La Laguna, Tenerife, Spain\\
$^{35}$ Facultad de Ciencias Astron\'omicas y Geof{\'\i}sicas -Universidad Nacional de La Plata. Paseo del Bosque S/N (1900). La Plata, Argentina\\
$^{36}$ CONICET, Rivadavia 1917, 1033, Buenos Aires, Argentina\\
$^{37}$ Instituto de Fisica, Universidad Nacional Aut\'onoma de M\'exico, Apdo. Postal 20-364, M\'exico\\
}

\date{\today}
\pagerange{\pageref{firstpage}--\pageref{lastpage}}

\topmargin-1cm
\maketitle

\label{firstpage}

\clearpage

\begin{abstract}
  The Baryon Oscillation Spectroscopic Survey (BOSS), part of the
  Sloan Digital Sky Survey (SDSS) III project, has provided the
  largest survey of galaxy redshifts available to date, in terms of
  both the number of galaxy redshifts measured by a single survey, and
  the effective cosmological volume covered. Key to analysing the
  clustering of these data to provide cosmological measurements is
  understanding the detailed properties of this sample. Potential
  issues include variations in the target catalogue caused by changes
  either in the targeting algorithm or properties of the data used,
  the pattern of spectroscopic observations, the spatial distribution
  of targets for which redshifts were not obtained, and variations in
  the target sky density due to observational systematics. We document
  here the target selection algorithms used to create the galaxy
  samples that comprise BOSS. We also present the algorithms used to
  create large scale structure catalogues for the final Data Release
  (DR12) samples and the associated random catalogues that quantify the
  survey mask. The algorithms are an evolution of those used by the
  BOSS team to construct catalogues from earlier data, and have been
  designed to accurately quantify the galaxy sample. The code used,
  designated {\sc mksample}, is released with this paper.
\end{abstract}

\section{Introduction}  \label{sec:intro}
The size of galaxy redshift surveys has grown exponentially over the last decade and will continue do so into the next, thanks to the continuing development of instrumentation to undertake mulit-object spectroscopy (MOS) on dedicated telescopes. The scientific driver for this dramatic increase is that galaxy redshift surveys provide a wealth of cosmological and extra-galactic information. The most easily accessible cosmological information is encoded in 2-point clustering statistics of the over-density field, which contain both the Baryon Acoustic Oscillation (BAO) and Redshift Space Distortion (RSD) signals. The BAO scale is a comoving large-scale enhancement in pairs of galaxies separated by $\sim$150\,Mpc, which can be used to track cosmological expansion. It arises from the propagation of sound waves in the early Universe \citep{Pee70,Sun70,Dor78}, and is quite insensitive to astrophysical processing that occurs on smaller scales; thus BAO experiments are affected by a low level of systematics (see review by \citealt{WeinbergReview} for a comparison of different methods). Redshift-Space Distortions arise from the peculiar velocities of galaxies within a comoving frame, which produce coherent distortions in the measured redshifts compared to those produced by the Hubble expansion \citep{Kai87}. As these velocities are gravitational in origin, the amplitude depends on the rate of structure growth, and hence RSD allow tests of General Relativity (GR) on large scales.

\begin{figure} 
\includegraphics[width=85mm]{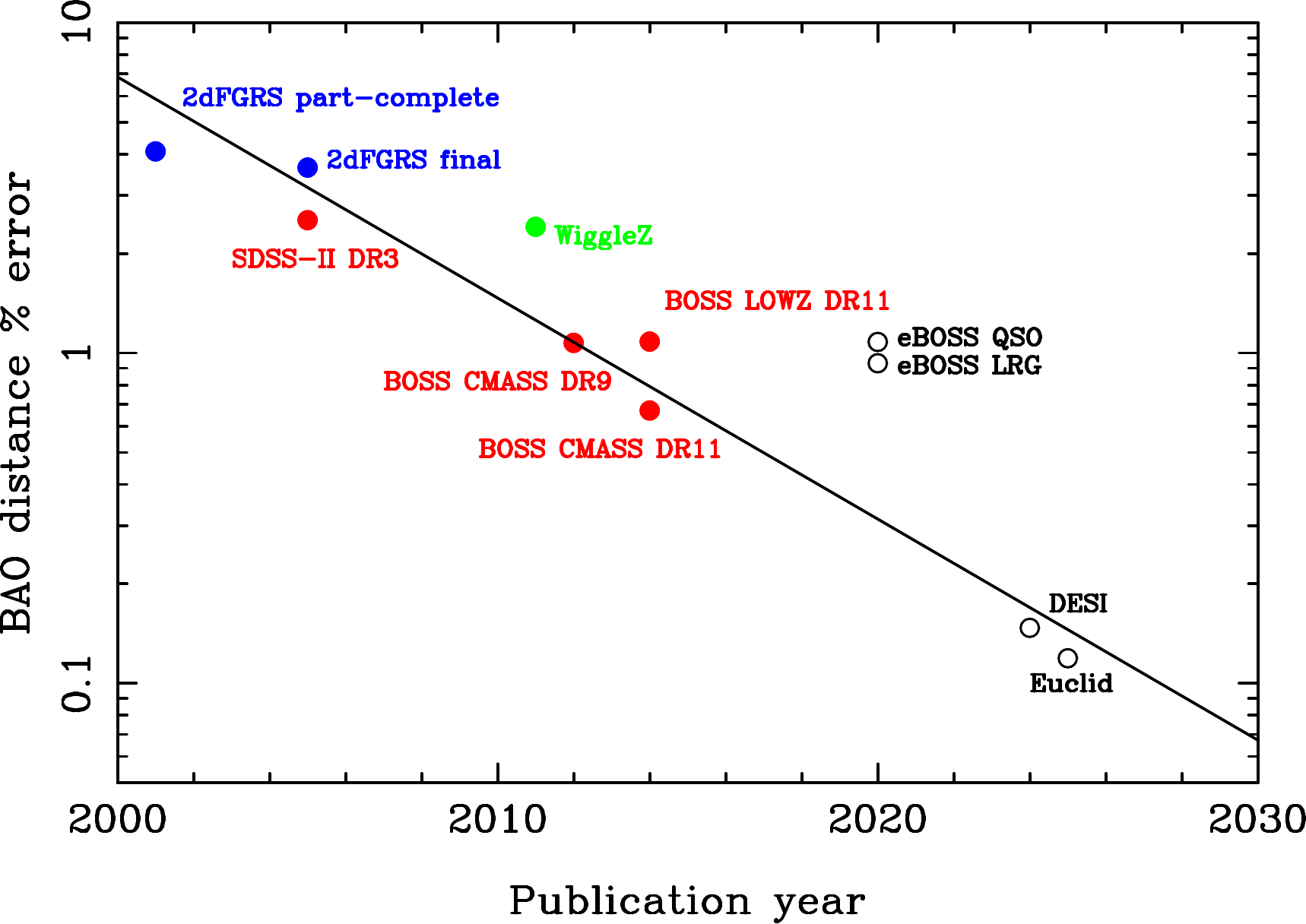} 
\caption{BAO measurement errors predicted for various surveys as a function of the year of publication. In order to calculate these with a consistent methodology we plot ``predictions'' using the code of \citet{seo07} based on a single number of galaxies, and volume for each survey. The surveys plotted are 2dFGRS early \citep{Per01} and final \citep{Col05}; SDSS-II LRGs \citep{Eis05}; WiggleZ \citep{blake11}; BOSS DR9 CMASS \citep{And12}; BOSS DR11 LOWZ \citep{Toj14} and CMASS \citep{And14}. In terms of survey volume, BOSS DR12 is very close to DR11 and we do not show it here. We also present approximate predictions for the eBOSS, DESI and Euclid future surveys (see text for details). \label{fig:surveys}}
\end{figure}

The BAO signature has now been detected in many different galaxy surveys and analysed using a variety of methods. To show the exponential growth in BAO measurements, Fig.~\ref{fig:surveys} presents the predicted error on the BAO scale expected for different surveys, calculated as if the clustering signal from different directions was optimally combined to provide the best possible single BAO position measurement. We include results from various stages of the 2-degree-Field Galaxy Redshift Survey (2dFGRS; \citealt{colless01,colless03}), Sloan Digital Sky Survey (SDSS; \citealt{Yor00}) and WiggleZ \citep{Drinkwater10}, and predictions for the continuation of the SDSS project with eBOSS \citep{Daw15}. For consistency, all calculations used the code of \citet{seo07}, approximating each survey as a single volume, limited in redshift and area, and sampled by a constant density of galaxies, with numbers approximately matching those of the actual surveys. Thus the results themselves are not precise and are designed to simply demonstrate the evolution rather than provide a quantitative comparison between experiments. The best fit line shows the growth in the impact of past surveys, following the development of Multi-Object Spectrographs (MOS) on the Anglo-Australian telescope \citep{Lewis98} and the Sloan telescope \citep{Gun06}, which continues to the next generation with a new MOS being developed for the Hobby-Eberly telescope (HETDEX; \citealt{Hill08}), the Mayall telescope (DESI; \citealt{levi13}), the VISTA telescope (4MOST; \citealt{deJong14}), the William Herschel Telescope (WEAVE; \citealt{Dal14}), the Subaru telescope (PFS; \citealt{Masahiro14}) and the satellite experiments Euclid \citep{laureijs11} and WFIRST \citep{Spergel15} . For clarity we only plot approximate DESI and Euclid predictions in Fig.~\ref{fig:surveys} to show the general expected trend from these new instruments, as our simplified approach is insufficient to provide a careful differential analysis of these future projects. Also, there is significant uncertainty in the predictions for Euclid, as a consequence of our lack of knowledge about the galaxy population targeted: the prediction here uses the predicted volume and galaxy density of \citet{laureijs11}. The higher redshift surveys of eBOSS and WiggleZ are inherently more difficult and consequently they lie above the line: they push into new redshift ranges, rather than to larger volumes. 

In this paper, we present the target selection and catalogue generation of the Data Release 12 (DR12; \citealt{DR12}) samples of galaxies selected from the Baryon Oscillation Spectroscopic Survey (BOSS; \citealt{Daw12}), which is part of SDSS-III \citep{Eis11}. The spectroscopic sample has two primary catalogues: LOWZ at $z \simlt 0.4$, and CMASS covering $0.4 \simlt z \simlt 0.7$ (see Section~\ref{sec:target} for details). An overview of the BOSS observations is provided in Section 2; see Dawson et al. (2012) for a full description of the survey.

The work presented here follows on from the analysis of previous data releases: DR12 is the third public SDSS data release containing BOSS spectroscopic results. The first was DR9 \citep{DR9}, when the survey was approximately one third complete. The creation of the large-scale structure catalogues from these data was outlined in \citet{And12}, alongside the isotropic BAO results, with the anisotropic results following in \citet{And13}. The development of a method to remove potential systematic errors in the clustering measurements caused by fluctuations in the target catalogue was presented in \citet{Ross12}. The DR9 catalogues were used extensively for science, which also tested the catalogues themselves. The clustering was compared with simulations in \citet{Nuza13}, and with full model fits in \citet{Sanchez12}. RSD were measured by \citet{Reid12} and enhanced by knowledge of passive galaxy evolution in \citet{Toj12}: the resulting GR tests were presented in \citet{Sam13}. Primordial non-Gaussianity was constrained by \citet{Ross13}, while \citet{Zhao13} reported neutrino masses, and \citet{Scoc13} examined the time variation of physical constants. This work led to further refinements of the catalogue creation algorithm for the analysis of the second public BOSS data release DR10 \citep{DR10}, which coincided with an internal release (called DR11). In particular, the code was rewritten into a modular version, called {\sc mksample}, new weights were used to correct for fluctuations in the expected target density, and new masks were used for ``bad'' areas. These refinements were presented alongside the BAO results for the CMASS sample in \citet{And14} and the LOWZ sample in \citet{Toj14}, and were confirmed to be robust to colour \citep{Ross14} and against possible systematics in the fit \citep{Var14}. As for DR9, the results were extensively used, further testing the catalogues: RSD measurements have been made in a number of different ways \citep{Beutler14,Sam14,Sanchez14,Chu13}, the bispectrum calculated and analysed \citep{Mar14a,Mar14b}, and neutrino mass constraints presented \citep{Beutler14b}. \citet{Saito15} account for redshift dependent selection effects and compare clustering and RSD with predictions from abundance matching. 


We have now analysed the final BOSS DR12 galaxy sample using an algorithm that builds on the work described above. This paper on the targeting algorithm and catalogue creation method is complemented by a series of papers measuring and analysing clustering, splitting the BOSS galaxies into sub-samples delineated by the primary targeting algorithms LOWZ and CMASS samples (see Section~\ref{sec:target} for details). BAO measurements are presented in configuration-space (Cuesta et al. 2015) and Fourier-space (Gil-Marin et al. 2015a), and RSD measurements made in Fourier-space are presented in Gil-Marin et al. (2015b). Two further support papers are provided in this set: Ross et al. (2015) considers the BOSS selection function in more detail, presenting the observational foot-print, masks for image quality and Galactic extinction, and weights to account for density relationships intrinsic to the imaging and spectroscopic portions of the survey. Vargas-Magana et al. (2015) presents systematic tests on the reconstruction algorithm used for anisotropic BAO analyses. A subsequent set of analyses to be released soon, will consider jointly analysing the full BOSS sample, without splitting by target selection.

Because the key cosmological measurements depend on the density field, galaxy properties (except how they trace this field, commonly quantified by a linear deterministic bias $b$), are unimportant once redshifts have been measured, and cosmological surveys are free to choose which galaxies to observe to optimise survey efficiency and the optimal bias $b$. BOSS targets luminous galaxies for spectroscopic observations as they have a large bias, are relatively easy to target, and have strong spectral features that ease redshift determination. The target selection adopted by BOSS is an extension of the targeting algorithms for the SDSS-II \citep{Eis01} and 2SLAQ \citep{Can06} Luminous Red Galaxies (LRGs), targeting fainter and bluer galaxies in order to achieve the desired number density of $\sim3\times10^{-4}$\,h$^3$Mpc$^{-3}$. The majority of the galaxies are old stellar systems whose prominent 4000\,\AA\ break makes them relatively easy to target using multi-colour data. The data from which the samples are targeted is described in Section~\ref{sec:data}, and the LOWZ and CMASS target selection algorithms are discussed in detail in Section~\ref{sec:target}.

In order to do large-scale structure analyses with the sample of spectroscopically observed galaxies, we have put together catalogues including information on the detailed angular and radial mask of the sample including the redshift completeness, the observing conditions when the imaging and spectroscopic observations were made, and the appropriate weights to give each object, as well as random (i.e., unclustered) catalogues with the same selection function.   These collectively make up the large-scale structure catalogues, whose contents are detailed in this paper.  

Key to creating these catalogues for the BOSS galaxy surveys is the ability to predict where we could have observed galaxies, as well as where galaxies exist, thus defining the survey or sample mask. This mask is intricately linked with the selection of galaxies: in general, corrections for selection effects can be applied to either the mask or the galaxy sample to produce a match between the two. In order to understand the mask, we need to understand both the target sample and the subsequent spectroscopy and redshift measurement, which we briefly summarise in Section~\ref{sec:observations}. The BOSS galaxy mask is quantified using a “random catalog”, a Poisson sampling of the volume covered by the selected galaxies, including any variations in density other than the cosmological clustering signal we wish to measure. The “3D mask” does not have to be quantified by a Poisson sampling, but this is a straightforward approach to this - in effect providing a Monte-Carlo sampling of the volume covered. This weighted random sample and the weighted galaxy sample form the starting point for the key BOSS galaxy clustering analyses. Section~\ref{sec:catalog} presents the method adopted by the BOSS team to prepare catalogues of galaxies and randoms, using routines made publicly available in a code called {\sc mksample}. This is a further extension of the code used for the early DR9 analyses, which is described in \citet{And12}.

Although the targeting algorithm adopted for BOSS is isotropic and the catalogue of target objects covers an angular area larger than that of spectroscopic observations, the mask is complicated by various anisotropic effects including variations in imaging depth due to recalibration of the SDSS photometric scale and rereduction of the imaging as the spectrosocpic survey progressed, variation of seeing, variation with stellar density caused by occultation by stars, the inability to measure spectra for close to another target observed at the same time, and the failure to measure spectra as a function of signal-to-noise ratio in the spectrum. These effects are often corrected by applying a weight to the galaxies (e.g. \citealt{Ross12}), but could instead be incorporated into the mask. The quality of the DR12 data is such that we can now observe systematic effects that couple radial and angular fluctuations, and we introduce 3D corrections for these. The manner adopted to deal with these effects for BOSS is described in Section~\ref{sec:systematics}.

BOSS includes a number of galaxy catalogues with different selection functions, some of which spatially overlap. The combination of these to optimally quantify the underlying matter overdensity field is non-trivial, and we present the method adopted by the BOSS team in Section~\ref{sec:combined}.

The {\sc mksample} code will be released upon publication of this paper, and we will also publish the resulting Large-Scale Structure catalogues, with a full datamodel describing each. These will all be linked from the main SDSS web site {\tt http://www.sdss.org/surveys/boss} .

\section{Data}  \label{sec:data}
\subsection{Imaging Data}  \label{sec:imdata}

The Sloan Digital Sky Survey (SDSS-I/II; \citealt{Yor00}) imaged
approximately 7,606\,deg$^2$ of the Northern Galactic Hemisphere and
600\,deg$^2$ of the Southern Galactic Hemisphere in the $ugriz$ bands
\citep{Fuk96,Smi02,Doi10}, using a specially designed camera
\citep{Gun98} on the 2.5m Sloan telescope \citep{Gun06} at the Apache
Point Observatory in New Mexico. The SDSS-III project \citep{Eis11}
obtained additional imaging to make the region of the Southern
Galactic Hemisphere contiguous, covering 3,172 deg$^2$. As part of
this effort, the original SDSS-I/II data and the SDSS-III data were
reduced with the latest versions of the SDSS image processing and
calibration pipelines \citep{Photo,Pie03,Pad08}. These data were
released as part of Data Release 8 \citep{DR8}, and form the parent
imaging catalogue for the BOSS galaxy target selection. There are a
number of differences between the processing performed for DR8
(see the DR8 paper \citealt{DR8} for a detailed discussion) and
earlier reductions; reproducing BOSS galaxy samples derived from the
imaging data requires using the appropriate algorithms.

\begin{table*}
\begin{tabularx}{\textwidth}{|l|X|}
  \hline
  {\bf spherical polygon} & The base unit of a {\sc Mangle\/} mask.  Spherical polygons are used to represent the boundaries of the imaging survey from which the targets are drawn, the circular fields defined by spectroscopic tiles, as well as regions to be removed from the survey footprint (e.g. the centerpost of each spectroscopic tile; see Sec.~\ref{sec:vetomasks} for a full list).\\
  \hline
  {\bf spectroscopic tile} & Output of the tiling algorithm providing a central location on the sky and list of targets to be observed for spectroscopic observations.  Each tile has a circular field-of-view of radius 1.49 degrees, and can be observed by multiple plates.\\
  \hline
  {\bf plate} & Physical plate with a hole drilled for each target, based on the anticipated airmass of observation.  Spectroscopic tiles may be observed using multiple plates.\\
  \hline
  {\bf chunk} & Basic unit of sky input to the tiling algorithm. It consists of a set of rectangles in a spherical coordinate system.  The SDSS-III BOSS survey is composed of 38 chunks.\\
  \hline
  {\bf sector} & The union of spherical polygons defined by a unique intersection of spectroscopic tiles.  The survey completeness is treated as uniform within a sector.\\
  \hline
\end{tabularx}
\caption{Basic definitions for the geometric description of SDSS-III BOSS observations and the LSS {\sc Mangle\/} masks.}
\label{tab:mangle}
\end{table*}
 
BOSS obtained spectra and redshifts for 1,372,737 galaxies over
9,376\,deg$^2$. The targets are assigned to tiles of diameter 3\,deg,
using a tiling algorithm that is adaptive to the density of targets on
the sky \citep{Tiling}. Spectra are then obtained using the BOSS
spectrographs \citep{Smee13}. Each observation is performed in a
series of 900\,sec exposures, integrating until a minimum
signal-to-noise ratio is achieved for the faint galaxy
targets. Redshifts are then measured using the methods described in
\citet{Bolton12}. The spectroscopic observations were split into
distinct areas of sky, which we call chunks, targeted separately and
sequentially in time, each defined by a subset of the total
footprint. Later chunks can overlap earlier chunks and recover
unobserved targets. The angular distribution of chunks 2-11, which are
special as they reflect early versions of the target selection (see
Appendix A), but also serve to show how the survey is built up from
chunks are shown in Fig.~\ref{fig:chunks}, and basic definitions for
geometrical descriptors used in this paper are provided in
Table~\ref{tab:mangle}.

The start of spectroscopic observations preceded the finalisation of
the DR8 imaging reductions, so the imaging data used by BOSS are based
on the photometric measurements {\em available at the time of tiling}
(see Section~\ref{sec:tiling}), which may differ from the quantities
available for an object in the DR8 catalog. BOSS targeting was
performed using three different versions of the reduction software
that resolves the catalogues from overlapping imaging data ({\sc
  RESOLVE}; see \citealt{DR8}). Chunks 1--4 used a version of the
{\sc RESOLVE} software tagged on 14-06-2009, chunks 5--11 used a version
tagged on 16-11-2009, and chunks 12 onwards used the same version as
that used to produce DR8. In total, 17\% of targets were targeted with
pre-DR8 {\sc RESOLVE} versions. Because these different versions of the
software selected different imaging data to be designated as
``primary'' (i.e. either the only or the best observation of this
object; see the DR8 documentation for more details), approximately 9\%
of the imaging data used for targeting CMASS galaxies is now
designated as secondary\footnote{i.e., there is an overlapping
  observation with higher quality photometry} in the DR8 database.

\subsection{Parent catalog}

The selection of galaxy targets for spectroscopic observation is based
on a parent catalogue of photometrically identified objects within the
imaging data.  The parent catalogue was based on objects chosen from
3172\,deg$^2$ in the Southern Galactic Cap (SGC) and 7606\,deg$^2$ in
the Northern Galactic Cap (NGC), as described in this section. The
SDSS imaging pipeline returns a number of different measurements of
the photometry of galaxies. Full descriptions may be found on the SDSS
website\footnote{\texttt{http://www.sdss3.org/dr8}} and in
\citet{Sto02}. For galaxy target selection we use three photometric
measurements, which have all been corrected for Galactic extinction
using the \citet{SFD98} dust maps.

The colours of galaxies are based on SDSS model magnitudes (denoted by
the subscript {\it mod}). These are determined by using the best-fit
(psf-convolved) deVaucouleurs or exponential profile fit in the $r$
band to determine the fluxes in the other bands (full details are
provided in \citealt{DR2}). Cuts in apparent magnitude are made with
``cmodel'' magnitudes (denoted by a subscript {\it cmod}). These are a
linear combination of the flux from the best fit exponential and
deVaucouleurs profile fit in each band separately\footnote{contrasted
  with model magnitudes where the fit in the $r$ band is used}
\begin{equation}
f_{\rm cmod} = (1 - P) f_{\rm exp} + P f_{\rm deV} \, , 
\end{equation}
where $P$ is the best-fit coefficient obtained from a fit of the
linear combination of the deVaucouleurs and exponential profile fits
to the image, and weights the different contributions (reported as
${\rm frac}_{\rm deV}$ by the SDSS pipelines), and $f$ represents the
flux ({\it not magnitude}) assuming an exponential or deVaucouleur
profile.  Star-galaxy separation compares the PSF magnitudes of
galaxies (denoted by a subscript {\it PSF}) with model or cmodel
magnitudes; PSF magnitudes underestimate the flux from extended
sources compared with the model fits (see \S~\ref{sec:separation} for
details). Finally, we use ``fiber2'' (denoted by subscript {\it fib2})
magnitudes to estimate the expected flux through the SDSS-III 2$"$
fibres.

The parent sample for the BOSS galaxy target selection is constructed
by selecting all detected objects that the photometric pipeline
classifies as galaxies, and that are chosen by {\sc RESOLVE} to be
``primary''. The targeting software uses the photometry of the primary
objects to select targets for spectroscopic follow-up. The variation
in selected targets from different imaging data is consistent with
that expected given the photometric uncertainties, and so we treat the
regions targeted with pre- and final DR8 imaging as statistically
identical.  We do not make any cuts on photometricity at this stage;
unphotometric data is discarded at the catalogue creation stage (see
\S~\ref{sec:vetomasks}). Users constructing their own samples for
science analyses are advised to use the CALIB\_STATUS flag to cut on
photometricity (restricting to photometric observations corresponds to
\texttt{CALIB\_STATUS==1}). We cull objects with suspect photometry as
reported in the flags set by the imaging pipeline. In particular, we
require objects that are detected in the $r$ and $i$ bands. In the
Image Processing pipeline, this is indicated by having one of the
BINNED1, BINNED2 or BINNED4 flags set in both the $r$ and $i$
bands. We also require that the OBJC\_FLAG flag, which is a
combination of the per-filter flags appropriate for the whole object
(the full definition is provided in \citealt{Sto02}) has
\begin{enumerate}
  \item Objects not to be saturated :\texttt{(NOT SATUR) OR (SATUR AND (NOT
  SATUR\_CENTER))},
  \item Blended objects : \texttt{(NOT BLENDED) OR (NOT NODEBLEND)},
  \item Other photometric quality flags : \texttt{(NOT BRIGHT) AND (NOT
  TOO\_MANY\_PEAKS) AND (NOT PEAKCENTER) AND (NOT NOTCHECKED) AND (NOT
  NOPROFILE)}.
\end{enumerate}

\section{Target selection}  \label{sec:target}
We now turn to the specifics of the target selection algorithms used
to define the BOSS spectroscopic galaxy samples.  We first summarize
the criteria that we wish our algorithm to satisfy
(\S~\ref{sec:reqs}), with the aim of defining a uniformly selected
sample over a broad redshift range. The galaxy sample is targeted
using two different algorithms, which we term ``LOWZ'' (detailed in
\S~\ref{sec:lowz}) and ``CMASS'' (for ``Constant (stellar) Mass'',
\S~\ref{sec:cmass}), respectively.  Star-galaxy separation is treated
differently in the CMASS sample than elsewhere in SDSS, as we describe
in \S~\ref{sec:separation}.  A variant of the CMASS algorithm was used
to explore the colour boundaries of the sample
(\S~\ref{sec:cmass_sparse}).

\subsection{Requirements and Criteria}
\label{sec:reqs}

The BOSS sample was designed to measure the BAO signature in the
two-point galaxy clustering signal, 
and in particular to meet error requirements on the measurement of the
angular diameter distance $d_A$ and Hubble parameter $H$ at $z=0.35$
and $z=0.6$. These
requirements can be met by a survey covering an area of approximately
10,000 deg$^2$ with a comoving number density of galaxies of $3\times
10^{-4}$\,h$^3$Mpc$^{-3}$ for $0.1 < z < 0.6$.  This density is close
to optimal for large-scale cosmological studies (e.g., \citealt{Kaiser86}). To efficiently undertake
such a survey using the Sloan telescope and spectrographs, we need to
select a sub-sample of the parent catalogue of photometrically
identified objects that fulfil the following criteria simultaneously:  
\begin{enumerate}
\item galaxies that lie in the desired redshift range $0.1 < z < 0.6$,
\item sufficient galaxies to meet the desired density over the full
  redshift range,
\item well-defined limits in stellar populations, to isolate a
  strongly clustered subsample of galaxies, 
\item \label{item:zmeas} redshifts that can be measured in a relatively short exposure
  with our telescope, 
\item few contaminating objects that are not part of the desired sample,
\item selectable uniformly across the desired area,
\item selection is not sensitive to systematic errors in the data used.
\end{enumerate}
The challenge of target selection is to provide an algorithm for
selecting the subsample of the parent imaging catalogue that optimally
meets these goals. Selection based solely on an apparent magnitude
cut, as used for the SDSS-I and -II Main Galaxy Sample \citep{Str02}
in general selects too many low redshift and low luminosity galaxies.
Rather, in BOSS we follow a similar philosophy to the selection of
Luminous Red Galaxies (LRGs) in SDSS-I and -II \citep{Eis01} and the
2SLAQ survey \citep{Can06} using colour-magnitude and colour-colour
cuts, selecting luminous galaxies with strong spectral features
(item~\ref{item:zmeas} above).

\begin{figure}
\includegraphics[width=1.0\columnwidth]{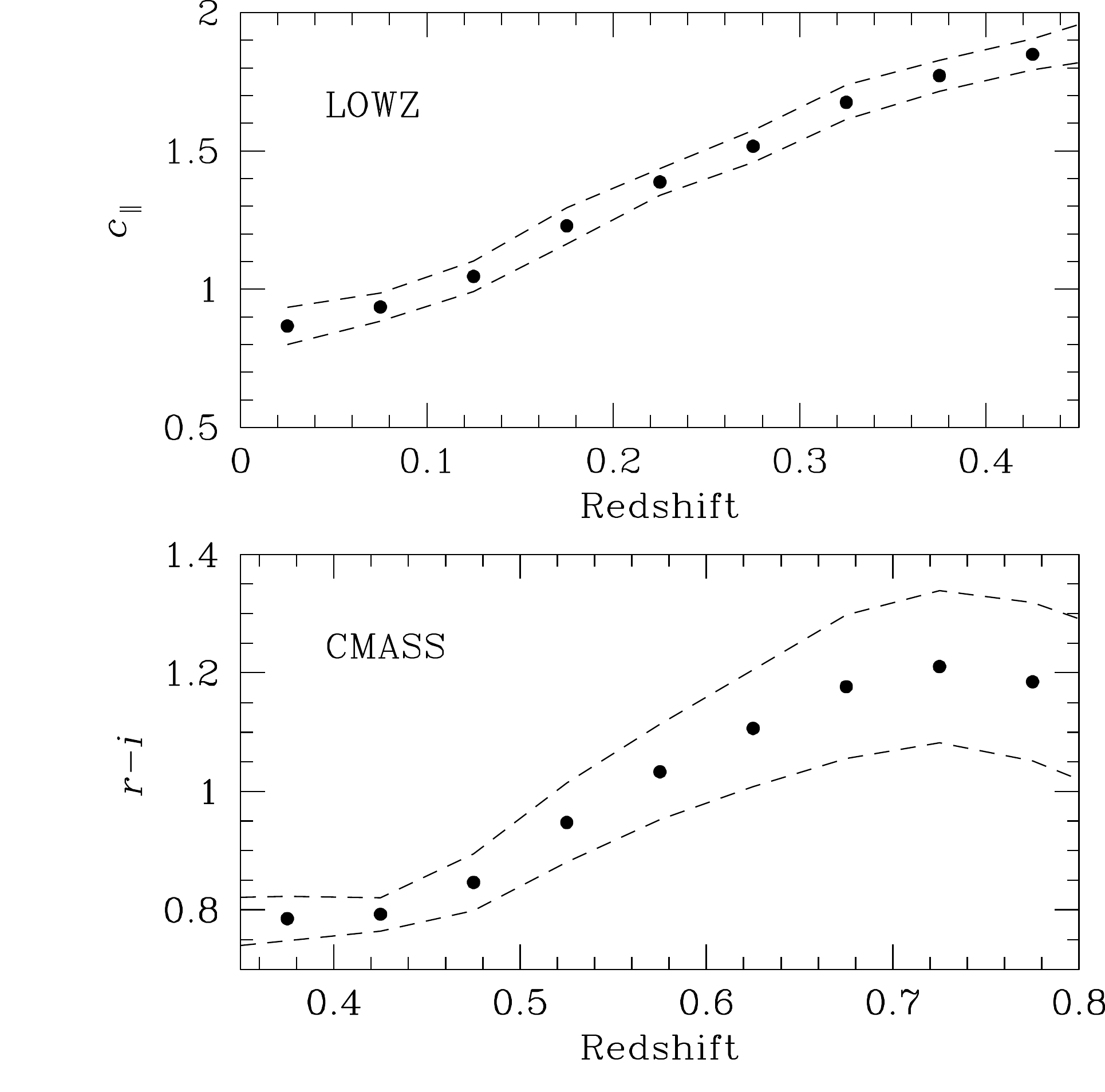}
\caption{Top panel: Black dots show median $\cll$ for LOWZ
  spectroscopically confirmed galaxies as a function of measured
  redshift, with the dashed lines showing the interquartile range. The
  efficiency of using this quantity to track redshift is clear. Bottom
  panel: Median $\rmod - \imod$ as a function of redshift for
  confirmed CMASS galaxies, with interquartile range (dashed
  lines). The way in which we can track the high-redshift locus of
  galaxies using this colour, and select as a function of redshift, is
  clear.}
\label{fig:colorredshift}
\end{figure}

At redshifts $z<0.4$, we can select such a sample by extending to
fainter LRGs than observed in SDSS-I and -II. At higher redshifts, we
do not restrict ourselves to red galaxies, and instead select an
approximately stellar mass-limited sample of objects of all intrinsic
colours.  As in \citet{Eis01}, two sets of colours are necessary to
describe the colour locus: one when the 4000\,\AA\ break lies in the
SDSS $g$-band, and the other when it redshifts into the $r$-band at
$z\sim0.4$. Selecting these two subsamples requires defining fiducial
colours that track the locus of a passively evolving population of
galaxies in $gri$ colour space. Following \citet{Eis01} and
\citet{Can06}, we define
\begin{eqnarray}
\cll & = &  0.7 (\gmod - \rmod) + 1.2(\rmod - \imod - 0.18)  \\
\cpp & = & (\rmod - \imod) - (\gmod - \rmod)/4.0 - 0.18 
\end{eqnarray}
to describe the low redshift locus and 
\begin{eqnarray}
\dpp & = & (\rmod - \imod) - (\gmod - \rmod)/8 \,,
\end{eqnarray}
to describe the high-redshift locus. As discussed above, the colours
are defined using SDSS model magnitudes, and are corrected for Milky
Way extinction. The efficiency of these selections to select luminous
galaxies as a function of redshift is demonstrated in
Fig.~\ref{fig:colorredshift}, which shows how $\cll$ and $\rmod -
\imod$ versus redshift for observed BOSS galaxies.

Where the targeting algorithms use colour selection, they are built on
model magnitudes, which are based on the flux measured through
equivalent apertures in all band and thus provide unbiased colours of
galaxies. Brightness limits are instead based on cmodel magnitudes,
which provide better estimates of the total light observed.

\begin{figure}
\includegraphics[width=1.0\columnwidth]{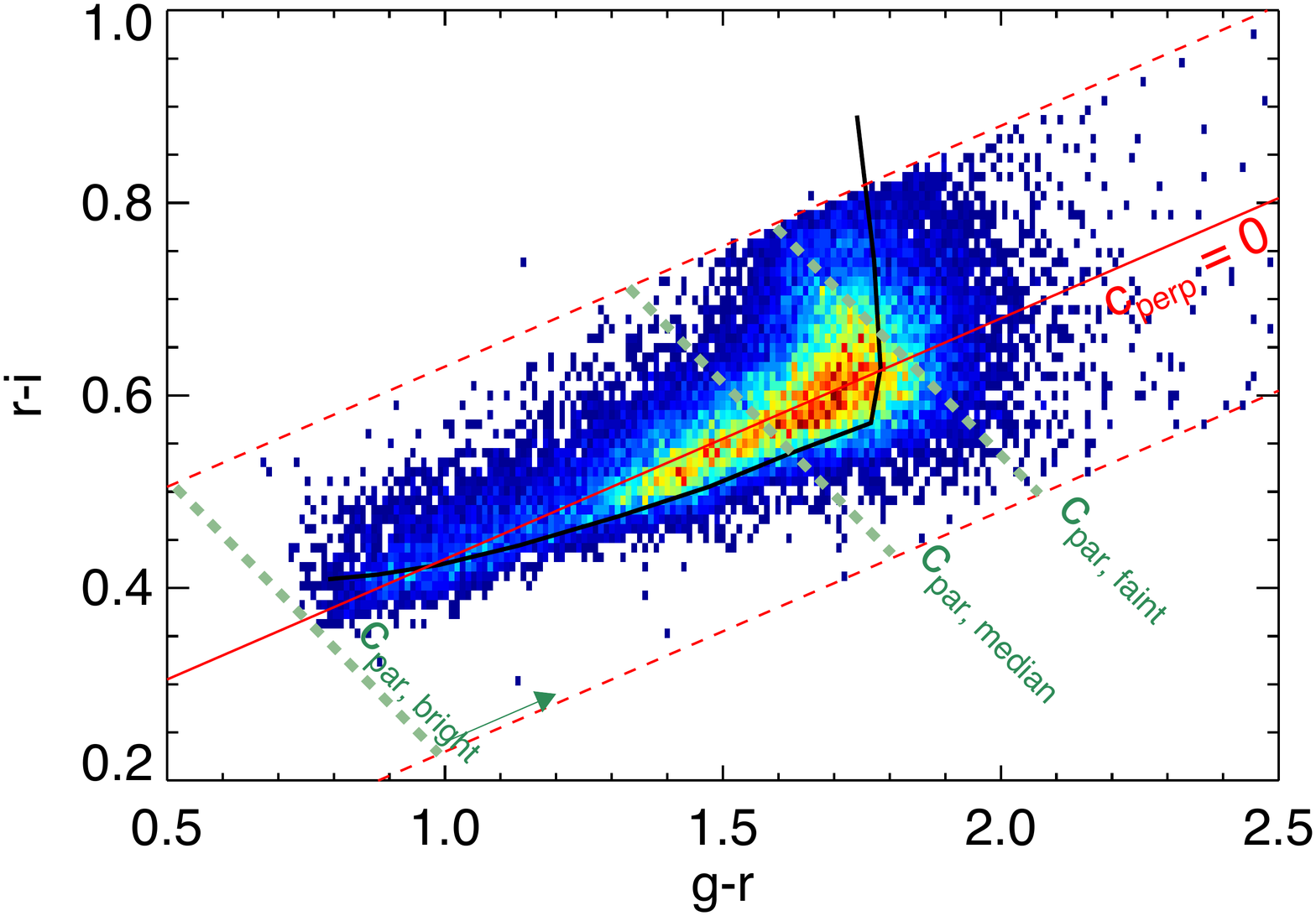}
\caption{Density plot of LOWZ galaxies in the $(g-r, r-i)$ colour
  plane; red corresponds to higher density and dark blue to lower
  density, in an arbitrary normalisation and linear scale. Redshift
  increases rightwards and upwards along the galaxies locus, starting
  at $z\simeq0.1$ on the bottom left corner. The knee on the galaxy
  locus is caused by the 4000\,\AA\ break transitioning between the
  $g-$ and $r$-band filters, and happens at $z\approx 0.4$. The
  colours $\cpp$ and $\cll$ are simple rotations of this colour plane,
  and trace the position of a target in parallel and perpendicular,
  respectively, to the data locus. The black thick line represents the
  passively evolving LRG model of \citet{Mar09}. The green and red
  dashed lines are the colour and magnitude targeting cuts -- see the
  main text for details. The few targets seen outside of the selection
  cut are due to differences in the targeting and final photometry,
  see Section~\ref{sec:data}. }
\label{fig:lowz_TS}
\end{figure}

\subsection{The LOWZ sample}\label{sec:lowz}

The LOWZ sample is designed to extend the SDSS-I/II Cut I LRG sample
\citep{Eis01} to $z \approx 0.4$ to fainter luminosities, in order to
increase the number density of the sample by roughly a factor of 3.
Fig.~\ref{fig:lowz_TS} shows how the colours $\cll$ and $\cpp$
describe the evolution of a passively evolving stellar population with
redshift. Redshift increases from the bottom left to upper right.  The
black line shows the passively evolving LRG model of \citet{Mar09}. The
\citet{Mar09} 'LRG' template is a model of a metal-rich population in
passive evolution containing a small fraction of a metal-poor coeval
population. This model was found to be a good fit to the
$g,r,i$~colours of luminous red galaxies (LRGs) from the 2SLAQ survey
\citep{Can06} as a function of redshift, over models containing star
formation in various amount. The same model also better fit the
overall luminosity evolution of BOSS galaxies \citep{Montero15}. The
knee seen in the galaxy locus corresponds to the transition of the
4000\,\AA\ break from the $g-$ to the $r-$band. The parameter $\cll$
quantifies the position of a galaxy along the main locus, and $\cpp$
characterises the departure of a galaxy from the centre of the locus;
$\cpp = 0$ lies approximately at the centre of the galaxy
distribution.

We select targets at low redshift ($z < 0.4$) around the predicted colour locus using
\begin{equation}
  |\cpp|  <  0.2,
\end{equation}
(red dashed lines in Fig.~\ref{fig:lowz_TS}) and we select the
brightest and reddest objects at each redshift using a sliding
colour-magnitude cut with $\cpp$ (an effective proxy for a photometric
redshift):
\begin{equation}
  \rcmod   <  13.5 + \cll/0.3.
\end{equation}
The dashed green lines in Fig.~\ref{fig:lowz_TS} show the effective
cuts in $\cll$ for three different $r-$band magnitudes: $r=16, 18.73$
and $19.6$ mag corresponding to the faint boundary, the median
magnitude and the bright boundary of the sample respectively. Thus
fainter objects must be redder to pass the cut.  This cut is the most
important criterion in the selection of LOWZ galaxies - it drives the
number density of the sample by effectively setting the magnitude
limit as a function of redshift, and aims to produce a constant number
density over the desired redshift range. The number of galaxies in the
sample is therefore highly sensitive to this cut (see
\citealt{Ross12,Toj14}). The resulting space density of the sample is
shown in Fig.~\ref{fig:nofz}; the sample is close to volume-limited
(constant space density at $\sim 3 \times 10^{-4}\,h^3\,\rm Mpc^{-3}$)
over the redshift range $0.2 < z < 0.4$.

We impose brightness limits on the targets, such that
\begin{equation}
16  < \rcmod < 19.6.
\end{equation}
The faint limit ensures a high redshift success rate. The bright limit
excludes a significant number of low-redshift blue galaxies that would
otherwise pass the colour cut, but also excludes a fraction of
brightest cluster galaxies in low-redshift massive clusters
\citep{Hoshino15}.  A bright cut was not needed in SDSS-I/II as such
galaxies were already targeted by the SDSS-I/II Main Galaxy Sample
\citep{Str02}, but a significant fraction of the BOSS footprint lies
outside that of SDSS-I and -II.

The star-galaxy separation follows the same procedure introduced in
\citet{Eis01} for the LRGs,
\begin{equation}
\rpsf - \rcmod  >  0.3\,.
\end{equation}
The cmodel magnitude is a proxy for a ``total'' magnitude for a
galaxy, while the PSF magnitude fits the unresolved component of the
object.  The difference between the two is therefore a measure of the
extendedness of the galaxy.

In summary, the LOWZ selection algorithm, as implemented after
commissioning, is as follows:
\begin{eqnarray}
\rcmod  & < & 13.5 + \cll/0.3 \\ 
|\cpp| & < & 0.2 \\
16 & < \rcmod < & 19.6 \\
\rpsf - \rcmod & > & 0.3 \label{eq:sgLOWZr}\,\,,
\end{eqnarray}

The galaxies in the LOWZ sample may be selected from the DR12 database
using the following flags, whose definitions can be found on the SDSS
website\footnote{\tt http://www.sdss.org}:
\begin{itemize}
\item BOSS\_TARGET1 \&\&  $2^0$ \qquad Objects targeted by the LOWZ algorithm.
\item SPECPRIMARY ==  1 \qquad Objects with spectra, removing
  duplicate observations. 
\item ZWARNING\_NOQSO  ==  0 \qquad Objects whose spectroscopic
  redshifts are cleanly measured.
\item CLASS\_NOQSO == 'GALAXY' \qquad Objects whose spectra are those
  of a galaxy (as opposed to a quasar or star). 
\end{itemize}

The basic properties of the LOWZ sample are presented in
\citet{Par13}, who fitted the small-scale clustering of the galaxies
using halo occupation distribution (HOD) modelling. They demonstrated
that these galaxies lie in massive haloes, with a mean halo mass of
$5.2\times10^{13}$\, $h^{-1}$M$_\odot$, a large-scale bias of
$\sim2.0$ and a satellite fraction of $12\pm2$\%. These galaxies
occupy haloes with average masses between those of the CMASS sample
and the original SDSS I/II LRG sample.

\subsubsection{Exceptions to the LOWZ targeting}

During the first nine months of BOSS observations, the incorrect
star-galaxy separation criterion was used to identify LOWZ targets,
removing a significant fraction of galaxies (see
Appendix~\ref{app:lowze}). To select a uniformly-targeted sample from
all LOWZ redshifts, with the selection criteria described in this
section, the simplest procedure is to avoid those data with the use of
an additional cut
\begin{itemize}
  \item TILEID$\ge 10324$,
\end{itemize}
where TILEID identifies spectroscopic tiles, and this cut corresponds
to chunk numbers larger than 6.

Further details on this issue, and other slight changes in the
targeting of LOWZ galaxies in early chunks, can be found in
Appendix~\ref{app:lowze}. Briefly, LOWZ targets in chunk 2, and LOWZ
targets in chunks 3-6, were selected with different algorithms from
those of subsequent data. For the purposes of a large-scale structure
catalog, in previous data releases we simply removed chunks 2-6 from
the LOWZ sample and the corresponding mask.  In \S~\ref{sec:combined}
we construct separate samples using the chunk 2 (``LOWZE2'') and
chunk 3-6 (``LOWZE3'') selections, and combine all three LOWZ catalogues
with the CMASS samples to construct a single unified sample
appropriate for analyses restricted to large scales, such as BAO
fitting. The effects of these changes on the density of galaxies
measured as a function of redshift can be seen in Fig.~\ref{fig:nofz}.

\subsection{The CMASS sample}\label{sec:cmass}

\begin{figure}
\includegraphics[height=1.0\columnwidth, angle=90]{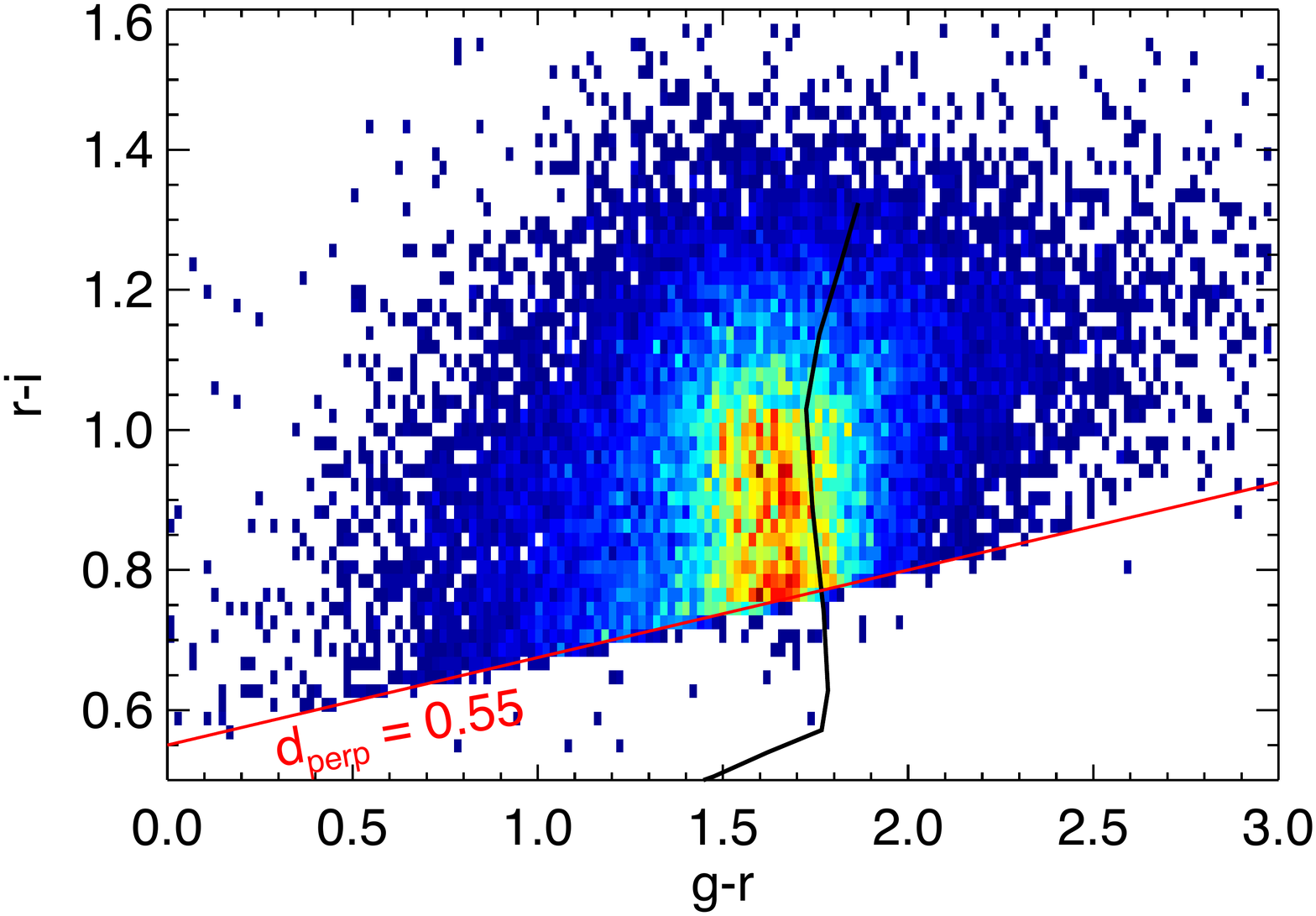}
\includegraphics[height=1.0\columnwidth, angle=90]{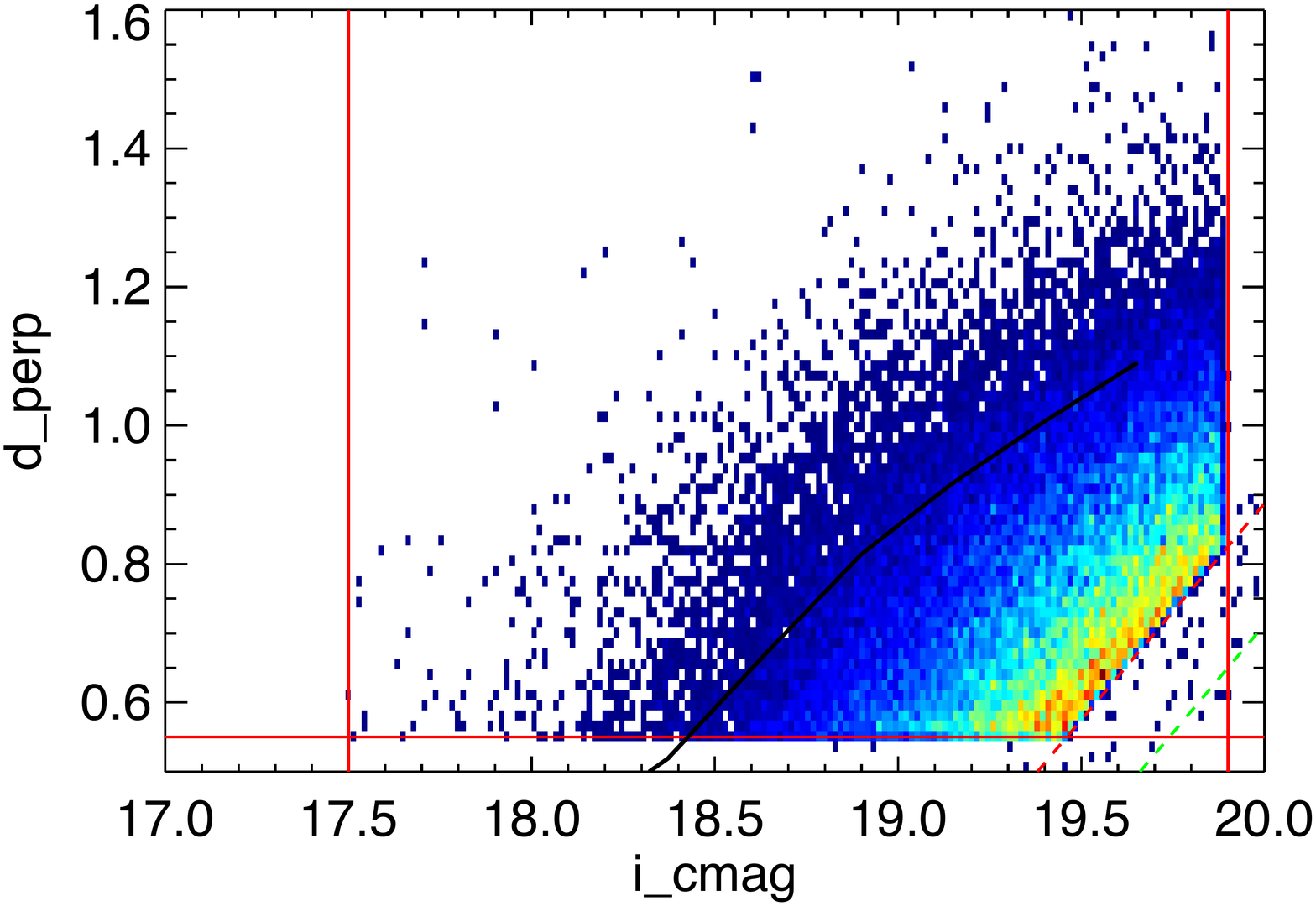}
\caption{Both panels show density plots of CMASS galaxies; red
  corresponds to higher density and dark blue to lower density in an
  arbitrary normalisation and linear scale. The black thick line shows
  the passively evolving LRG model of \citet{Mar09}. {\it Top:}
  redshift increases upwards, starting at $z\simeq0.4$ at $\dpp =
  0.55$. {\it Bottom:} the sliding cut in $\dpp$ with $i-$band
  magnitude, designed to select an approximately stellar-mass complete
  sample. Stellar mass increases with the perpendicular distance to
  the sliding cut, represented here by the red dashed line - see
  \citet{Mar13} for details. The green dashed line shows the sliding
  cut adopted for the CMASS SPARSE sample (see
  Section~\ref{sec:cmass_sparse}). Vertical solid lines show the
  magnitude limits. On both panels, the small fraction of targets that
  lie outside of the selection cut are due to differences in the
  targeting and final photometry, see Section~\ref{sec:data}. Only
  chunks greater than 6 are shown. }
\label{fig:cmass_TS}
\end{figure}

The CMASS sample uses similar selection cuts to those utilised by the
Cut-II LRGs from SDSS-I/II and the LRGs in 2SLAQ, but extends them
both bluer and fainter in order to increase the number density of
targets in the redshift range $0.4 < z < 0.7$ and get closer to a mass
limited sample.

The quantity $\dpp$ (Fig.~\ref{fig:cmass_TS}) effectively discards
low-redshift galaxies by choosing
\begin{equation}
  \dpp > 0.55.
\end{equation}
We do not apply any further colour cuts, with the exception of a
sliding colour-magnitude cut that selects the brightest objects at
each redshift, in such a way as to keep an approximately constant
stellar mass limit over the redshift range of CMASS according to the
passively evolving model of \citet{Mar09}:
\begin{equation}
  \icmod  <  min(19.86 + 1.6(\dpp - 0.8),19.9) \,\,.
\end{equation}

This approach is a significant departure from SDSS-I/II Cut-II and
2SLAQ LRGs - which consisted of essentially a flux-limited sample with
a colour cut to isolate the reddest galaxies. 

We impose model and magnitude limits as follows: 
\begin{eqnarray}
17.5 & < & \icmod <   19.9 \,\,, \\
\ifib & < & 21.5 \,\,. \\
\end{eqnarray}
The faint magnitude limits are set to ensure a high redshift success
rate, whereas the bright limit protects against some low-redshift
interlopers. In the first 14 tiling chunks, CMASS objects were
targeted with $\ifib < 21.7$, but the redshift failure rate at the
faint end of this range was quite poor, so we revised this limit to
the final value of $\ifib < 21.5$.

To exclude outliers with problematic deblending, we further impose the
following cuts on colour and $r_{dev,i}$ (the effective radius in the
fit to the deVaucouleurs profile for the $i$-band magnitude, measured
in pixels):
\begin{eqnarray}
\rmod - \imod & < & 2 \\
r_{dev,i} & < & 20.0\,{\rm pix} \,\,.
\end{eqnarray}
These cuts remove a very small fraction of targets. The CMASS
star-galaxy separation is described in detail in the next section.

CMASS galaxies can be selected from the DR12 database using the
following flags:
\begin{itemize}
\item BOSS\_TARGET1 \&\& $2^1$
\item SPECPRIMARY == 1
\item ZWARNING\_NOQSO == 0
\item CLASS\_NOQSO == 'GALAXY'
\end{itemize}

The basic clustering properties of the CMASS sample are presented in
\citet{Whi11}, which fitted the small-scale clustering of the galaxies
using HOD modelling. They showed that these galaxies lie in massive
haloes, with a mean halo mass of $2.6\times10^{13}$\,
$h^{-1}$M$_\odot$, a large-scale bias of $\sim2.0$ and a satellite
fraction of $10$\%. These galaxies occupy haloes with lower masses
than those of the LOWZ sample, although the bias is similar, a
consequence of them being at higher redshift.

CMASS galaxies are massive, with $M_* >10^{11} M_\odot$
(e.g. \citealt{Chen12, Mar13}), and the majority are dominated by old
stellar populations with low star-formation rates
(e.g. \citealt{Chen12, Thomas13, Toj12}). \citet{Mar13} argues that
the CMASS sample becomes significantly incomplete at stellar masses
$M_* < 10^{11.3} M_\odot$ and $z > 0.6$ for a Kroupa initial mass
function, and is roughly consistent with a volume-limited sample at
higher masses and lower redshift. \citet{Thomas13} presented similar
results, showing that stellar velocity dispersions of BOSS galaxies
peak at $\sim$240\,kms$^{−1}$ with a narrow distribution virtually
independent of redshift. Most recently, \citet{Leauthaud15} quantified
the stellar mass completeness of CMASS and LOWZ using data from the
Stripe 82 region of sky along the celestial equator - a narrow, but
deeper subset of the SDSS imaging survey region, that is 2 magnitudes
deeper than the single epoch SDSS imaging \citep{Annis14}. Using the
Stripe 82 Massive Galaxy Catalog \citep{Bundy15}, they estimate that
CMASS is 80\% complete at $\log_{10}(M_*/M_{\odot}) \geq 11.6$ in the
redshift range $z=[0.51,0.61]$. The stellar mass completeness of CMASS
decreases at lower and higher redshifts and the denomination
``constant mass'' should be considered only as a loose approximation
outside of the redshift window $z=[0.51,0.61]$. However, the
combination of LOWZ and CMASS yields a spectroscopic sample that is
80\% complete at $\log_{10}(M_*/M_{\odot}) \geq 11.6$ at
$z<0.61$. Compared to cut-II LRGs, CMASS galaxies have a larger range
of properties including morphology \citep{Masters11}, star-formation
rates \citep{Thomas13,Chen12} and star-formation histories
\citep{Toj12}, partly because no red cut has been imposed on the g-r
observed-frame colour. It should be noted however that, for example,
galaxies with detectable emission-lines (hence hosting very young
stellar populations) still represent only 4\% of the sample (see
\citealt{Thomas13}).

\subsubsection{Exceptions in CMASS targeting flag}\label{sec:CMASS_exceptions}

The meaning of BOSS\_TARGET1 \&\& $2^1$ (the CMASS targeting flag) evolved during the first 14 chunks of the survey. Therefore  BOSS\_TARGET1 \&\& $2^1$ will {\em not} select CMASS galaxies (as defined by the equations in the previous sections) in these regions, and further
subsampling is required based on galaxy colours and magnitudes to recover the final
selection in these regions. Alternatively, these chunks can be explicitly excluded. For the first 14 chunks the following exceptions should be noted:
\begin{itemize} 
\item {\bf Chunks 1 \& 2:} The data taken in the commissioning phase (chunks 1 \& 2) used a
significantly broader selection criteria (see Section~\ref{sec:CMASS_summary}), and therefore must be dealt
with carefully.  
\item {\bf Chunks 3-6}: The data taken in chunks 3-6 used a slightly looser $\icmod$ cut, selecting instead on
$\icmod<19.92+1.6(\dpp-0.8)$. 
\item {\bf Chunks 1-14}:  As mentioned above, the cut in $\ifib$ changed during the
survey. In chunks 1-14 the targeting required $\ifib < 21.7$. 
\end{itemize}
With the exception of chunk 1, all of these chunks are included in the LSS catalogue after applying the required subsampling based on colours and/or magnitudes.

\subsubsection{Star-Galaxy Separation in the CMASS sample}
\label{sec:separation}

The difference between psf and model magnitudes is a measure of the
extendedness of a source, thus making it useful to separate stars from
galaxies. For the commissioning phase of the survey we applied a
star-galaxy separation criterion identical to that used in the 2SLAQ
survey \citep{Can06}, a sloping cut in $\ipsf-\imod$:
\begin{equation}
\label{eq:sgCMASSi}
  \ipsf - \imod > 0.2 + 0.2(20 - \imod)\, .
\end{equation}
Whilst this cut is effective at removing the bulk of the stars,
roughly 6.9\% of $\sim$7000 CMASS targets from the commissioning runs
had stellar spectra, mainly cool M-dwarfs.

Fig.~\ref{fig:stargal} displays the distribution of the
spectroscopically classified stars (blue) and galaxies (red) in the
$\ipsf-\imod$ vs $\imod$ and $\zpsf-\zmod$ vs $\zmod$ planes for these
commissioning targets. As expected, the stars preferentially occupy
lower values of psf$-$model than the galaxies, and so applying a more
restrictive cut would remove more stars but at the expense of removing
some galaxies. For a maximum loss of just 1\% of galaxies we found the
linear cuts that would remove the largest numbers of stars. These
cuts, shown as the black lines in Figure \ref{fig:stargal}, remove
31\% and 52\% of the stars that remained in the commissioning data for
the $i-$ and $z-$band cuts respectively. Since the $z-$band cut,
\begin{equation}
\label{eq:sgCMASSz}
  \zpsf - \zmod  >  9.125 - 0.46 \zmod,
\end{equation}
performed significantly better, it was added to the original $i-$band
cut (Eq.~\ref{eq:sgCMASSi}) for all data from Chunk 3 onwards (i.e.,
after the commissioning runs), such that targets have to pass both
cuts to be selected. Even though the $z-$band cut alone removes the
vast majority of the stars excluded by the $i-$band cut, we kept the
$i-$band cut in place to ensure that we could apply a consistent
star-galaxy separation throughout the survey. This is achieved simply
by retroactively applying the $z-$band cut to the commissioning data.

\begin{figure}
\includegraphics[width=0.95\columnwidth]{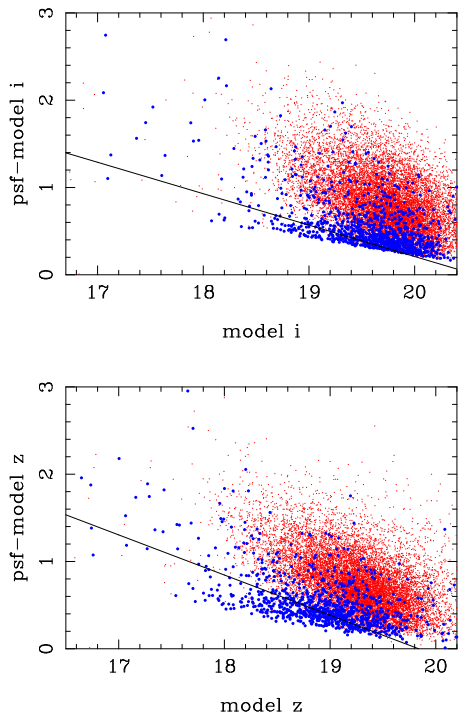}
\caption{The distribution of spectroscopically confirmed stars (large
  blue points) and galaxies (small red points) in the psf-model vs
  model $i$-band (top) and $z$-band (bottom) planes selected in the
  CMASS sample of the commissioning data. The black lines are the
  linear cuts that remove the most spectroscopically confirmed stars
  whilst removing less than 1\% of the galaxies. The $z-$band cut was
  added to the original $i-$band cut targeting from chunk 3 onwards.}
\label{fig:stargal}
\end{figure}

Since these star-galaxy separation criteria measure the compactness of
the objects in the SDSS imaging, their effectiveness will depend on
the imaging PSF. Based on the commissioning data,
Fig.~\ref{fig:stargal_psf} shows how the fraction of stars and
galaxies removed by the new $z-$band criteria depends on the $r-$band
PSF. The fraction of galaxies removed is fairly flat at $\sim$ 1\% for
PSF FWHM $<$ 1.5$"$ and then rapidly increases at higher FWHMs. The
fraction of stars removed displays the opposite
trend. Fig.~\ref{fig:stargal_psf} also presents the numbers of stars
and galaxies as a function of $r-$band FWHM, demonstrating that the
vast majority of the sample is selected from imaging with FWHM
$<$1.5$''$. This slight seeing-dependent star-galaxy separation will
result in the imprint of a spatial dependence in the density of
galaxies across the survey, which can be corrected using
seeing-dependent weights (see \ref{sec:sys} for details).

\begin{figure}
\includegraphics[width=0.95\columnwidth]{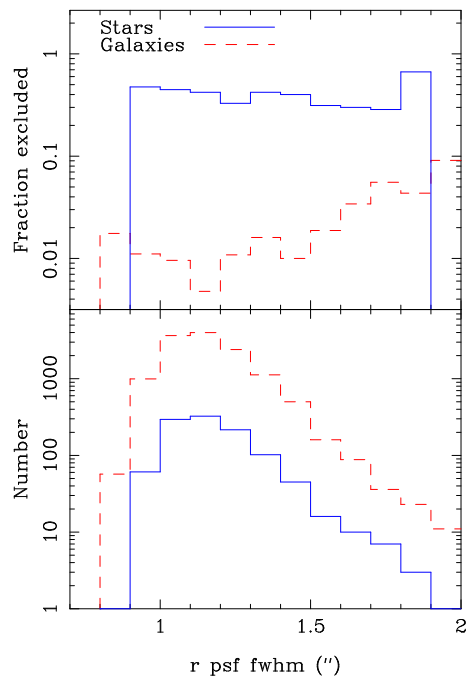}
\caption{The dependence of the star galaxy separation on the FWHM of
  the imaging PSF. The top panel shows the fraction, and the bottom
  panel the number, of spectroscopically classified stars and galaxies
  in the commissioning data that are excluded by the additional
  $z-$band star-galaxy separation as functions of $r-$band FWHM.}
\label{fig:stargal_psf}
\end{figure}

Whilst the above analysis addresses the fraction of galaxies lost due
to the addition of the $z-$band star galaxy separation criteria, it
provides no indication of how many compact galaxies were removed by
the original $i-$band cut. To investigate this issue we combined the
deep coadded SDSS Stripe82 imaging \citep{Abazajian09,Annis14} with
near-infrared $J$ and $K-$ band imaging from the UKIDSS Large Area
Survey Data Release 4 \citep{Lawrence07,Casali07,Hewett06,Hambly08} in
order to define a robust set of stars and galaxies over an area of 150
deg$^2$. A $J-K < 1.1$ colour cut provides an excellent separation
between stars and galaxies in the colour-magnitude region that the
CMASS galaxies occupy. When this information is combined with the
higher S/N measurement of $\zpsf-\zmod$ from the coadded imaging we
can confidently separate stars and galaxies. Using these data we
estimate that the final star-galaxy separation cuts removes 2.3\% of
the full sample of galaxies selected by the CMASS colour cuts.

\subsubsection{Summary of CMASS target selection}\label{sec:CMASS_summary}

In summary, the CMASS target selection for the bulk of the survey is as follows: 
\begin{eqnarray}
\icmod & < & 19.86 + 1.6(\dpp - 0.8) \\
17.5 & < \icmod <  & 19.9 \\
\dpp & > & 0.55 \\
\ipsf - \imod & > & 0.2 + 0.2(20 - \imod) \\
\zpsf - \zmod & > & 9.125 - 0.46 \zmod \\
\rmod - \imod & < & 2 \\
\ifib & < & 21.5 \\
r_{dev,i} & < & 20.0\,{\rm pix} \,\,.
\end{eqnarray}

During commissioning (chunks 1 and 2),
we used significantly looser criteria; the CMASS\_COMM sample
(BOSS\_TARGET\&\&$2^2$), just under 25000 galaxies, was selected as
follows:
\begin{eqnarray}
\icmod & < & 20.14 + 1.6(\dpp - 0.8) \\
17.5 & < \icmod <  & 20.0 \\
\dpp & > & 0.55 \\
\ipsf - \imod & > & 0.2 + 0.2(20 - \imod)  \\
\rmod - \imod & < & 2 \\
\ifib & < & 22 \\
\end{eqnarray}

See other exceptions to these criteria in Section~\ref{sec:CMASS_exceptions}.

\subsection{Sparse Sampling Cuts}\label{sec:cmass_sparse}

Motivated by the wish to study objects of slightly lower stellar mass
and bluer intrinsic colour, we designed the CMASS\_SPARSE sample. It
extends the CMASS selection by altering the $\icmod$-$\dpp$ sliding
colour-magnitude cut to
\begin{eqnarray}
\icmod & \ge & 19.86 + 1.6(\dpp - 0.8) \\
\icmod & < & 20.14 + 1.6(\dpp - 0.8) \,\,, 
\end{eqnarray}
with the other cuts unchanged (i.e., the area between the red and
green dashed lines in the bottom panel of
Fig.~\ref{fig:cmass_TS}). These galaxies were randomly subsampled down
to a number density on the sky of 5\,deg$^{-2}$, corresponding to
approximately 1 in 10 targets. This sample was selected across the
full BOSS footprint.

CMASS\_SPARSE galaxies may be selected with
\begin{itemize}
\item BOSS\_TARGET1 \&\& $2^3$
\item SPECPRIMARY == 1
\item ZWARNING\_NOQSO == 0
\item CLASS\_NOQSO == 'GALAXY'
\end{itemize}
after excluding the commissioning chunks.

Altering the CMASS target selection in this way produces a sample of
galaxies at somewhat lower redshift and stellar mass. The median
redshift of CMASS\_SPARSE is $z=0.51$, with a stellar mass
distribution that peaks at $10^{11.2}$ M$_\odot$ (using the stellar
masses of \citealt{Chen12}), relative to the peak CMASS mass of
$10^{11.4}$ M$_\odot$.

\section{Spectroscopic observations}  \label{sec:observations}
\subsection{Previously known redshifts}  \label{sec:known}

Fractions of the LOWZ and CMASS targets have a previous robust object
classification and redshift determined from the SDSS-II survey
\citep{Yor00,Abazajian09}. We therefore matched our target sample to a
sample of ``known objects'' with pre-determined secure classifications
and redshifts and did not spectroscopically reobserve these galaxies
within BOSS. This subsample of targets has a complicated angular
distribution on the sky: the majority of the NGC was covered by
SDSS-II, but only a few stripes in the SGC were observed.  These
pre-observed targets account for 43\% (9\%) of the LOWZ targets in the
north (south).  A much smaller fraction of CMASS targets were
pre-observed: 1.7\% (0.7\%) in the N (S).

\subsection{Target Collation and Spectroscopic Tiling}  \label{sec:tiling}

We start with the list of targets provided by the target selection
algorithms detailed above, and remove targets with known
redshifts as defined above.  The {\em tiling algorithm} assigns the
remaining targets to spectroscopic tiles.  The sky was tiled in a
piecemeal fashion as the survey progressed; each of these regions is
called a ``chunk''; see \S~\ref{sec:imdata} and \citet{Daw12} for further details.  DR12
contains observations from 38 chunks.  The survey mask and collated
target catalogue both indicate the chunk to which a region or specific
object was assigned.

The tiling algorithm \citep{Tiling} determines the location of the
$3^\circ$ diameter spectroscopic tiles and allocates the available
fibres among the targets, including targets from other programmes
within BOSS.  Because of the size of the cladding on the fibres,
fibres may not lie within 62" of one another on a given spectroscopic
tile.  The algorithm therefore divides target galaxies into
friends-of-friends groups with a linking length of 62$''$, and then
assigns fibres to the groups in a way that maximizes the number of
targets with fibres.  The choice of which galaxies are assigned fibres
is otherwise random.  The algorithm adapts to the density of targets
on the sky, such that regions with a larger than average number
density tend to be covered by more than one tile.  For the DR12
sample, 42\% (55\%) of the area in the north (south) is covered by
multiple tiles, and the number density of CMASS targets is larger by
4.7\% (3.4\%) in those regions.  The tile overlap - target density
correlation is less pronounced for the LOWZ sample (1.6\% and 2.4\%
enhancement in north and south, respectively).  The LOWZ sample
constitutes only 35\% of the galaxy targets, and particularly in the
north many galaxies in dense regions already have spectra from the
SDSS-II and thus were not targeted for SDSS-III BOSS spectroscopy (see
Sec.~\ref{sec:known}).

Fibre collisions are partially resolved only in the multiple tile
regions, and therefore may not be representative of the unresolved
fibre collisions in lower target density regions.  Fibre-collided
galaxies cannot simply be accounted for by reducing the completeness
of their sector, since they are a non-random subset of targets
(conditioned to have another target within 62$''$).  As discussed
further in Sec.~\ref{sec:FB}, we provide a set of weights that treat
these objects as if they were observed, and assign their weight to the
nearest object of the same target class.  Finally, since quasar
targets are given higher priority by the tiling algorithm, we account
for their presence by simply including a 62$''$ veto mask (see
Sec.~\ref{sec:vetomasks}) around each high priority quasar target.

\subsection{Spectroscopic Reductions}

\begin{figure} \includegraphics[width=85mm]{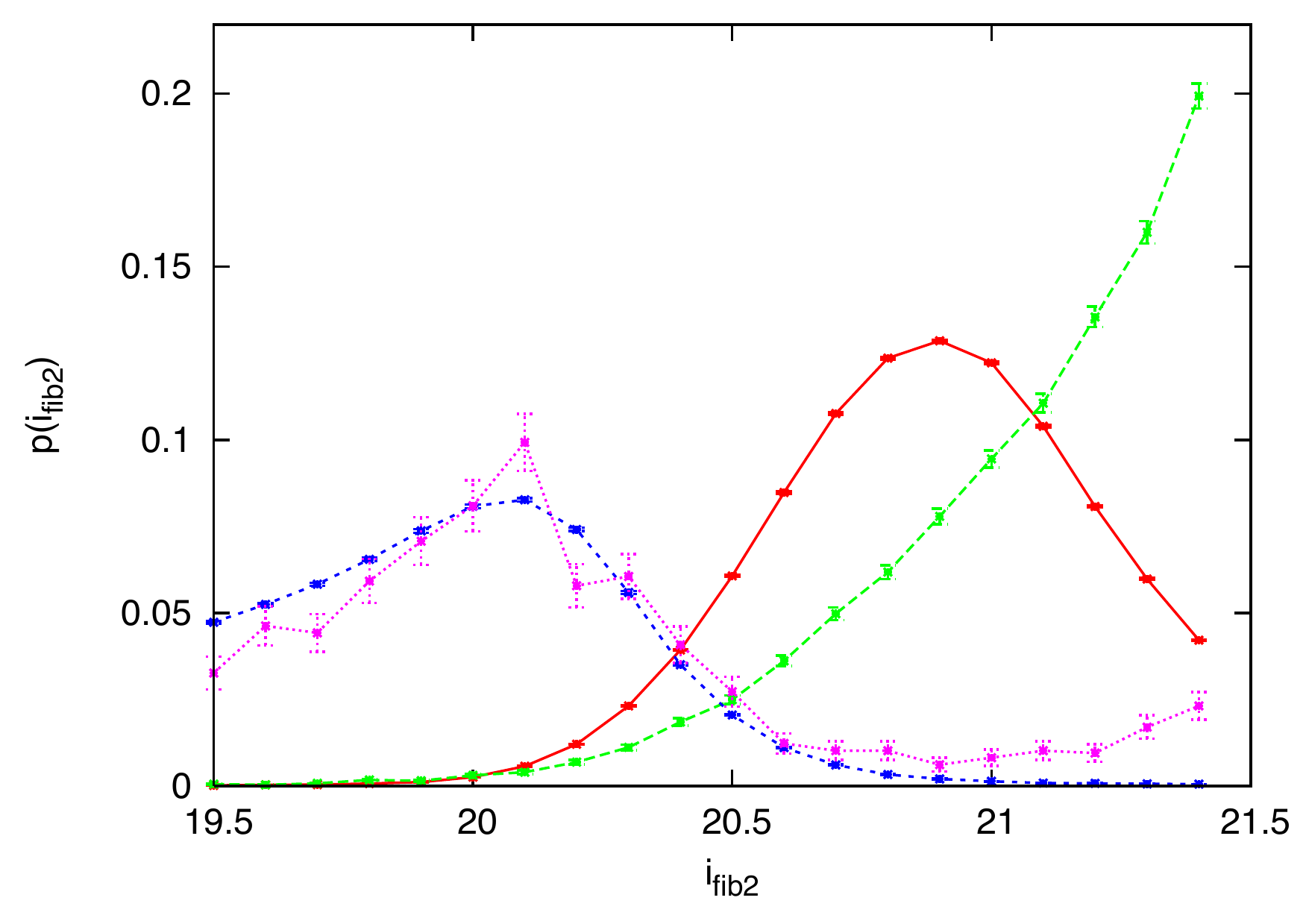} \caption{Normalised
    $i_{\rm fib2}$ distributions of redshift failures (green, dashed)
    and redshift successes (red, solid) for the CMASS sample. Redshift
    failures constitute 1.8\% of the CMASS targets observed by
    SDSS-III BOSS. These are contrasted against normalised
    distributions for the LOWZ sample of redshift failures (pink,
    dotted) and redshift successes (blue, dashed). Error bars were
    calculated assuming Poisson statistics. Note that some LOWZ galaxies
    have $i_{\rm fib2}<19.5$, which is why the normalisation for LOWZ
    curves looks lower than for CMASS.} \label{fig:zdist}
\end{figure}
Each ``tile'' output from the tiling algorithm specifies a central
location on the sky and the list of targets to be observed.  Physical
plates are drilled at the University of Washington based on the
anticipated airmass of observation.  Multiple plates can cover the
same tile, and plates may be observed on multiple nights until the
desired signal-to-noise ratio is reached \citep{Daw12}.

The BOSS spectroscopic reduction pipeline is detailed in
\citet{Bolton12}, with minor updates given in \citet{DR12}.  The final
DR12 catalogues used the v5\_7\_0 tag of the {\sc idlspec2d}
software package\footnote{\texttt{http://www.sdss3.org/svn/repo/idlspec2d/tags/v5\_7\_0/}}
for spectroscopic calibration, extraction, classification, and
redshift analysis.  We restrict the large-scale structure catalogues to
only include data from plates with PLATEQUALITY set to
``good.'' The criteria for this designation are a minimum of three
exposures, the number of spectroscopic pixels flagged as bad must be
less than 10\%, and a minimum signal-to-noise ratio requirement for both the
blue and red arms of the spectrograph must be met \citep{Daw12}.

The classification and redshift of each object are determined by a
Maximum Likelihood fit of the coadded spectra to a linear combination
of redshifted “eigenspectra” in combination with a low-order
polynomial.  The polynomial (quadratic for galaxies, quasars, and
cataclysmic variable stars; cubic for all other stars) allows for
residual extinction effects or broadband continua not otherwise
described by the templates. The templates are derived from a
rest-frame principal-component analysis (PCA) of training samples of
galaxies, quasars and stars using stellar population templates at the
BOSS resolution (from \citealt{Mar13}). The reduced $\chi^2$ versus
redshift is measured in redshift steps corresponding to the
logarithmic pixel scale of the spectra, where
$\Delta \log_{10}(\lambda) = 0.0001$.  Galaxy templates are fit from
$z=-0.01$ to $1.00$, quasar templates from $z=0.0033$ to $7.00$, and
star templates from $z=-0.004$ to $0.004$ ($\pm 1200\,$kms$^{-1}$).
The template fit with the best reduced $\chi^2$ is selected as the
classification and redshift, with warning flags set for poor
wavelength coverage, broken/dropped and sky-target fibres, and best
fits which are within $\Delta \chi^2/\rm{dof} = 0.01$ of the next best
fit (comparing only to fits with a velocity difference of more than
1000~kms$^{-1}$).  This method is a development of that used for the
SDSS DR8 \citep{DR8}, and is explained in further detail in
\citet{Bolton12}, and in \citet{DR9,DR10}.

For galaxy targets, a dominant source of false identifications is due
to quasar templates with unphysical fit parameters, {\em e.g.}, large
negative amplitudes causing a quasar template emission feature to fit
a galaxy absorption feature.  Thus, for galaxy targets, the best
classification and redshift are selected only from the fits to galaxy
and star templates, and we restrict the sample to fits the pipeline
classifies as robust. The results of these fits are tabulated in the
``*\_NOQSO'' versions of various quantities in the LSS catalogues.

Table~\ref{tab:basic_props} lists the total number of CMASS and LOWZ
targets that were assigned a fibre within the survey footprint
($\bar{N}_{\rm obs}$) as well as the breakdown for each of the three
possible outcomes: the number of CMASS and LOWZ targets robustly
classified as stars ($\bar{N}_{\rm star}$) or galaxies ($\bar{N}_{\rm
  gal}$), and the number of targets for which the pipeline failed to
find a robust classification and redshift ($\bar{N}_{\rm fail}$).  A
total of 2.3\% (3.4\%) of CMASS targets are stars and 1.6\% (2.1\%)
are redshift failures in the north (south).  Only 0.6\% of LOWZ
targets are stars and 0.5\% are redshift failures.

Fig.~\ref{fig:zdist} demonstrates that the pipeline is less likely to
obtain a successful redshift for CMASS targets with fainter $i_{\rm
  fib2}$ magnitudes.  Section~\ref{sec:zfail} discusses how we account
for this strong dependence in the redshift failure weights.

\section{Large scale structure catalogue creation}  \label{sec:catalog}
The creation of the BOSS large-scale structure catalogues involves a
number of steps.  We start with a list of targets based on the target
selection procedure described above, with the previously known
redshifts and outcome of the spectral analysis for each object for
which we have a spectrum, matched to this list. Next we construct the
survey mask, which specifies the regions of the sky that will be
included in the LSS catalogues and the completeness in each included
region. Finally, we use the mask and observed redshifts to generate a
set of ``random'' galaxies, Poisson sampling the sky coverage
specified by the mask with the same expected density distribution as
the galaxies.  The random galaxies are assigned redshifts to match the
distribution of the target sample.  Together, the data and random
catalogues can be used for statistical analyses such as $N$-point
functions. These steps and some of the subtleties involved are now
described in detail.

\subsection{Mask}  \label{sec:mask}

\begin{table*}
\begin{center}
\begin{tabular}{lrrrrrrrr}
Property & NGC & SGC & total  & NGC & SGC & total & NGC & NGC \\ \hline
Sample & & CMASS & & & LOWZ & & LOWZE2 & LOWZE3 \\ \hline
$\bar{N}_{\rm gal}$ &607,357 & 228,990 & 836,347 & 177,336 & 132,191 & 309,527 & 2,985 & 11,195 \\
$\bar{N}_{\rm known}$ &11,449 & 1,841 & 13,290 & 140,444 & 13,073 & 153,517 & 2,730 & 6,371 \\
$\bar{N}_{\rm star}$ &14,556 & 8,262 & 22,818 & 1,043 & 976 & 2,019 & 24 & 61 \\
$\bar{N}_{\rm fail}$ &10,188 & 5,157 & 15,345 & 868 & 602 & 1,470 & 21 & 55 \\
$\bar{N}_{\rm cp}$ &34,151 & 11,163 & 45,314 & 4,459 & 4,422 & 8,881 & 16 & 167 \\
$\bar{N}_{\rm missed}$ &7,997 & 3,488 & 11,485 & 10,295 & 3,499 & 13,794 & 114 & 609 \\
$\bar{N}_{\rm used}$ &568,776 & 208,426 & 777,202 & 248,237 & 113,525 & 361,762 & 4,336 & 15,380 \\
$\bar{N}_{\rm obs}$ &632,101 & 242,409 & 874,510 & 179,247 & 133,769 & 313,016 & 3,030 & 11,311 \\
$\bar{N}_{\rm targ}$ &685,698 & 258,901 & 944,599 & 334,445 & 154,763 & 489,208 & 5,890 & 18,458 \\
Total area (deg$^2$) &7,429 & 2,823 & 10,252 & 6,451 & 2,823 & 9,274 & 144 & 834 \\
Veto area (deg$^2$) &495 & 263 & 759 & 431 & 264 & 695 & 10 & 55 \\
Used area (deg$^2$) &6,934 & 2,560 & 9,493 & 6,020 & 2,559 & 8,579 & 134 & 779 \\
Effective area (deg$^2$) &6,851 & 2,525 & 9,376 & 5,836 & 2,501 & 8,337 & 131 & 755 \\
Targets / deg$^2$ & 98.9& 101.1& 99.5& 55.6& 60.5& 57.0& 43.4& 23.5\\
\end{tabular}
\end{center}
\caption{Basic parameters of the DR12 CMASS,LOWZ, LOWZE2, and LOWZE3 samples.  We track these classifications on a sector-by-sector basis in order to compute the BOSS fibre completeness in each sector of the survey.  In this table we report $\bar{N}_{X} = \sum_{sectors} N_X$, the sum over all sectors retained in the final BOSS mask.  Target classification counts and areas for the LOWZE2 and LOWZE3 samples are reported for chunk 2 and chunk 3-6, respectively.  To estimate the target density for those samples, we use the full NGC footprint to reduce cosmic variance.}
\label{tab:basic_props}
\end{table*}

We use the {\sc Mangle\/} software \citep{Mangle} to track the areas
covered by the BOSS survey and the angular completeness of each
distinct region; our terminology is summarised in
Table~\ref{tab:mangle}.  The mask is constructed of spherical
polygons, which form the base unit for the geometrical decomposition
of the sky.  The angular mask of the survey is formed from the
intersection of the imaging boundaries (expressed as a set of
polygons) and the spectroscopic tiles.  We define each unique
intersection of spectroscopic tiles to be a sector
\citep[see][]{Tiling,Teg04,DR8}.

We compute sector completeness based on the distribution of targets
across various outcomes of the tiling pipeline and spectroscopic
reductions.  In each sector (indexed by $i$) included in the large
scale structure catalog, we distinguish the following outcomes
(separately for each target class):
\begin{enumerate}
  \item galaxies with redshifts from good BOSS spectra (we denote the
    number in each sector by $N_{\rm gal,i}$),
  \item galaxies with redshifts from pre-BOSS spectra ($N_{\rm known,i}$),
  \item spectroscopically-confirmed stars ($N_{\rm star,i}$),
  \item objects with BOSS spectra from which stellar classification or
    redshift determination failed ($N_{\rm fail,i}$),
  \item objects with no spectra, in a fibre collision group with at
    least one object of the same target class ($N_{\rm
      cp,i}$), \footnote{{\it cp} is used because each galaxy exists in
      a ``close-pair'' with another}
  \item objects with no spectra, if in a fibre collision group then with no
    other objects from the same target class ($N_{\rm missed,i}$).
\end{enumerate}
These quantities, summed over all sectors included in the LSS
catalogues, are given in Table~\ref{tab:basic_props}.  As each target
is classed by one of these descriptors, we have that the total
number of targets in sector $i$ is
\begin{equation}  \label{eq:ntarg}
  N_{\rm targ,i} = N_{\rm star,i}+N_{\rm gal,i}+N_{\rm fail,i}+
                 N_{\rm cp,i}+N_{\rm missed,i}+N_{\rm known,i},
\end{equation}
and we define the number of targets observed by BOSS as
\begin{equation}
  N_{\rm obs,i} = N_{\rm star,i}+N_{\rm gal,i}+N_{\rm fail,i}.
\end{equation}
Matching our analyses for DR9, DR10 and DR11, the LOWZ catalogue is
then cut to $0.15<z<0.43$, and the CMASS catalogue is cut to
$0.43<z<0.7$ to avoid overlap, and to make the samples
independent. The number of galaxies used in the final catalogue
$N_{\rm used}$ is the subset of $N_{\rm gal,i}+N_{\rm known,i}$ that
pass these redshift cuts.

\begin{figure*}
 \includegraphics[width=85mm]{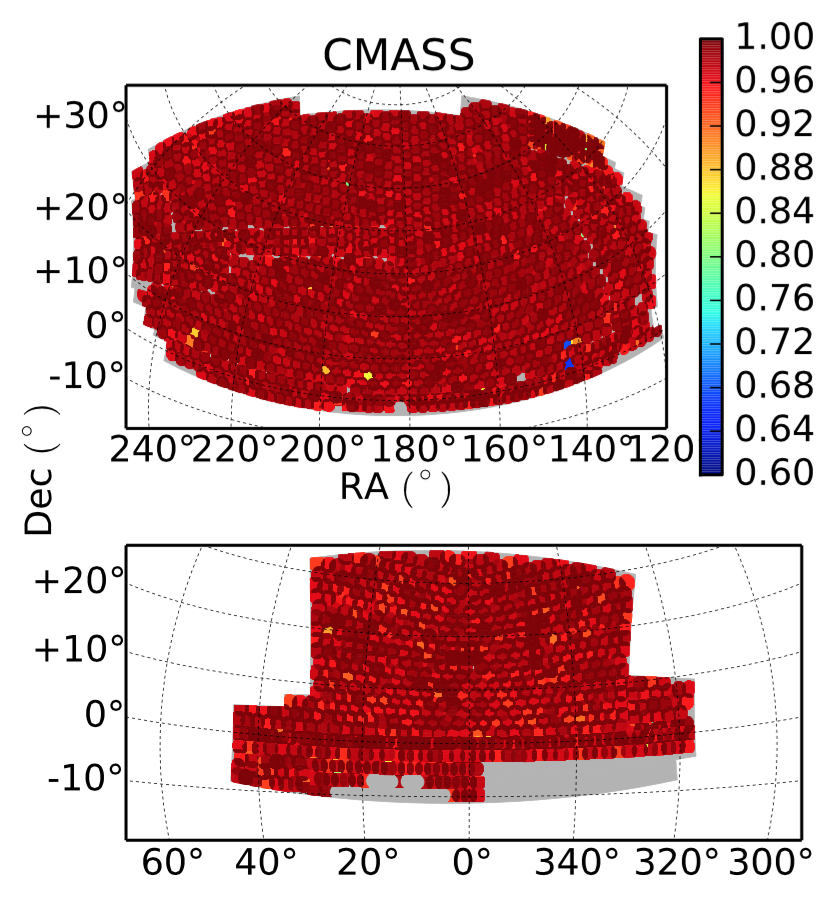} 
 \includegraphics[width=85mm]{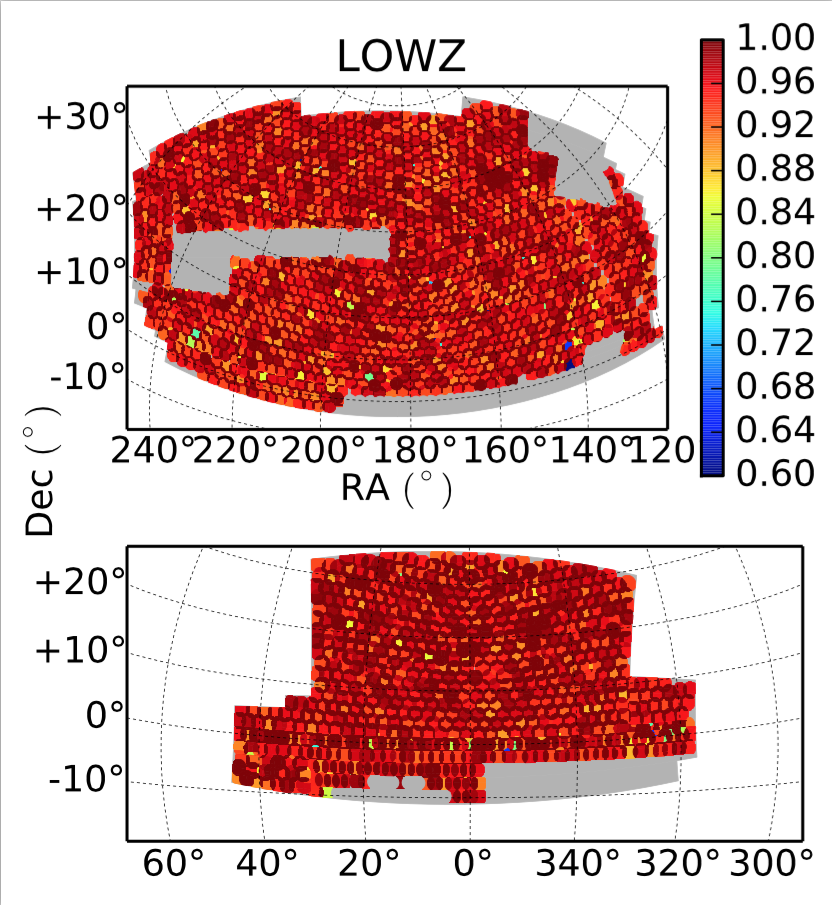} 
 \caption{Completeness maps for both the LOWZ and CMASS samples in the
   north and south Galactic caps. The mean completeness is 98.8\% for
   the CMASS sample shown in the left panels, and 97.2\% for the LOWZ
   sample in the right-hand panels. Gaps correspond to early chunks as
   shown in Fig.~\ref{fig:chunks}. Each patch of different colour
   corresponds to a plate, with the colour determined by the
   completeness of that plate. This is surrounded by the higher
   completeness regions that overlap that plate with other
   plates. This leaves a pattern that looks like a darker,
   higher-completeness “mesh”, covering the survey.}
\label{fig:mask}
\end{figure*}

From these descriptions, we define a BOSS fibre completeness in sector $i$
\begin{equation}  \label{eq:comp}
  C_{\rm BOSS,i} = \frac{N_{\rm obs,i}+N_{\rm cp,i}}{N_{\rm star,i}+N_{\rm gal,i}+N_{\rm fail,i}+N_{\rm cp,i}+N_{\rm missed,i}}.
\end{equation}
This completeness definition excludes the ``known'' objects
observed by SDSS-II.  $C_{\rm BOSS,i}$, shown in Fig.~\ref{fig:mask},
is recorded in the mangle mask files released with the LSS catalogues
and is used in the random catalogue generation (see
Sec.~\ref{sec:rancat}).  By this definition, the area-weighted average
completeness is 99\% (97\%) for the CMASS (LOWZ) samples.  We compute
the effective mask area in Table \ref{tab:basic_props} by weighting
the used area of each sector by its completeness.

The boundaries of the spectroscopic tiles can be seen by eye in
Fig.~\ref{fig:mask} as discontinuities in the value of completeness;
the unique intersection of those tiles define individual sectors, in
which we treat the BOSS fiber completeness as uniform. On average, the
completeness is larger in regions covered by more than one
spectroscopic tile.  The raw sky area covered by spectroscopic tiles
is 10338\,deg$^2$, of which 10252\,deg$^2$ remain (7429\,deg$^2$ in
the NGC and 2823\,deg$^2$ in the SGC) after restricting the mask to
sectors for which every planned tile has been observed with ``good''
\texttt{PLATEQUALITY}.

\begin{figure}
\includegraphics[width=1.0\columnwidth]{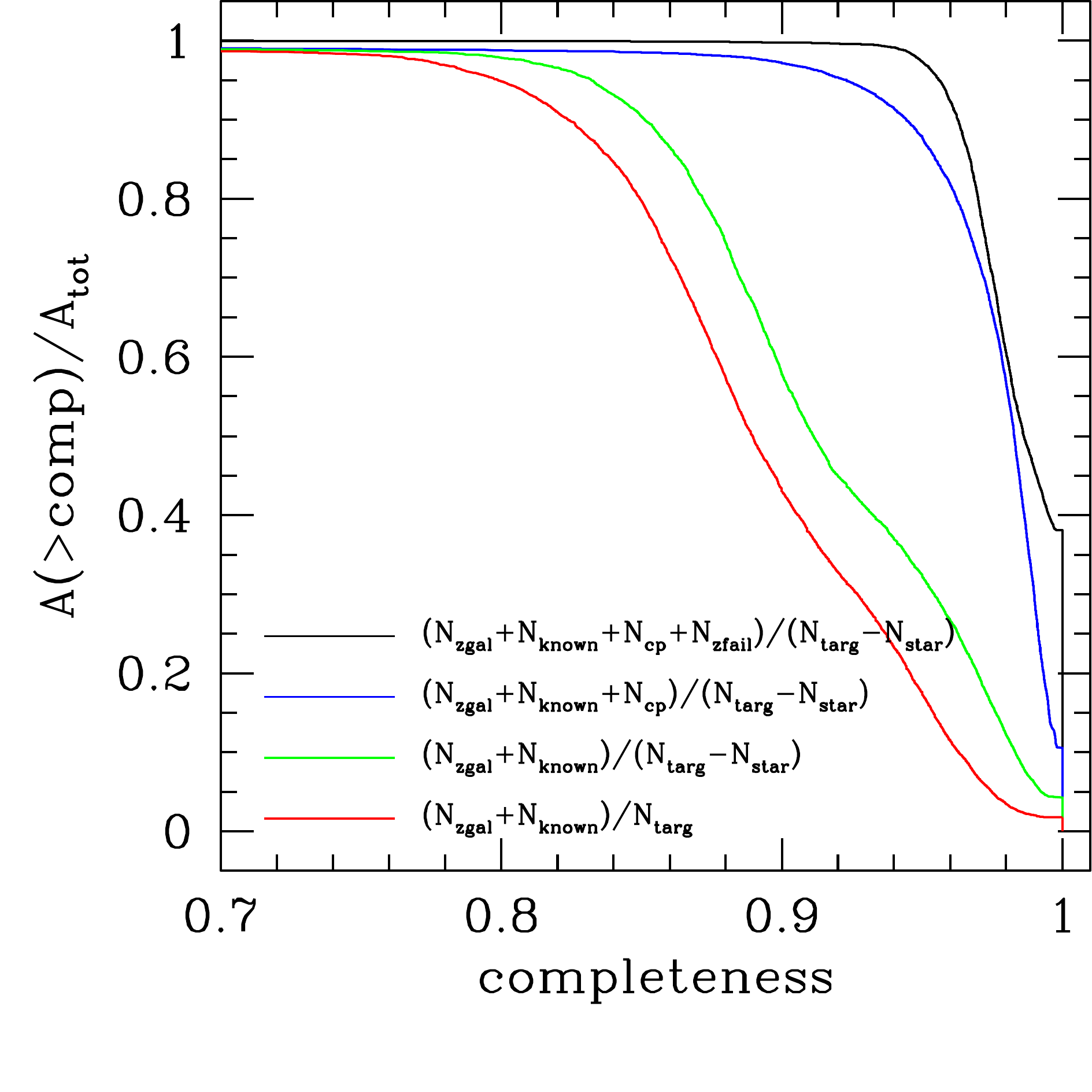}
\caption{The fraction of the total survey area that has a target
  completeness greater than the value shown, where target completeness
  is defined as the number of good galaxies spectroscopically observed
  in BOSS and those with previously known redshifts divided by the
  number of targets calculated in each sector as
  $C_{\rm targ,i} = (N_{\rm gal,i}+N_{\rm known,i})/N_{\rm targ,i}$,
  as in Eq.~(\ref{eq:targ_comp}).  We compare this completeness with
  those we would have obtained had we not had to include various
  classes of targets. If there had been no stars in our target list,
  the completeness would have been
  $(N_{\rm gal,i}+N_{\rm known,i})/(N_{\rm targ,i}-N_{\rm star,i})$
  (green line). If additionally we had not had to deal with fibre
  collisions, we would have observed a completeness
  $(N_{\rm gal,i}+N_{\rm known,i}+N_{\rm cp,i})/(N_{\rm targ,i}-N_{\rm
    star,i})$
  (blue line), and if additionally there were no redshift failures
  $(N_{\rm gal,i}+N_{\rm known,i}+N_{\rm cp,i}+N_{\rm fail,i})/(N_{\rm
    targ,i}-N_{\rm star,i})$
  (black line). From the definition of $N_{\rm targ,i}$ in
  Eq.~(\ref{eq:ntarg}) we see that the remaining decrement of the
  black line from $C_{\rm targ,i}=1$ is due to missed galaxies
  $N_{\rm missed,i}$.}
\label{fig:cumu}
\end{figure}

We also define a galaxy redshift completeness, assuming that stars are
always correctly classified spectroscopically
\begin{equation}  \label{eq:red_comp}
  C_{\rm red,i} = \frac{N_{\rm gal,i}}{N_{\rm obs,i}-N_{\rm star,i}},
\end{equation}
and define a target completeness 
\begin{equation}  \label{eq:targ_comp}
  C_{\rm targ,i} = \frac{N_{\rm gal,i}+N_{\rm known,i}}{N_{\rm targ,i}},
\end{equation}
which gives the number of good galaxies spectroscopically observed in
BOSS combined with previously known redshifts divided by the number of
targets calculated in each sector. Fig.~\ref{fig:cumu} shows the
fraction of the total BOSS area that has target completeness greater
than a specified value, and how this would change if we coud ignore
various effects. This shows the relative importance of different
categories of targets to the target completness of BOSS, from the
least important, which is redshift failures, to fibre collisions,
which is the most important.

Previous LSS catalogues (DR9, DR10, DR11) had to deal with sizeable
regions where BOSS spectra were not complete, and we made a
number of cuts on sectors to include in the LSS catalogues to minimise
the impact of this effect. In particular, sectors meeting any of the
following criteria were removed from the LSS mask:
\begin{itemize}
\item $C_{\rm BOSS,i} < 0.7$ (Eqn.~\ref{eq:comp}); removing
  part-complete sectors on the edges of the survey missing a
  significant fraction of redshifts.
\item $C_{\rm red,i} < 0.8$ (Eqn.~\ref{eq:red_comp}) and $N_{\rm
    gal,i} > 10$; removing regions with bad spectroscopic
  observations.
\item $N_{\rm obs,i} = 0$ and there is not another sector within 2\,$^\circ$
  in the $\pm$ right ascension or declination directions; removing isolated regions
  without galaxies.
\end{itemize}

But this was not done for the DR12 sample. If we had additionally
applied the fibre completeness cut (first criterion above), for DR12
we would have rejected an additional 30 (56)\,deg$^2$ from the CMASS
(LOWZ) mask; if instead we had applied the redshift success cut in
DR12 (second criterion above), we would have rejected an additional
1.7 (1.4)\,deg$^2$ from the CMASS (LOWZ) mask.  The difference between
the earlier mask selection and the algorithm described above applied
to DR12 constitute negligible changes on the survey mask.  The two
algorithms agree to within 0.3\% of the total mask area for both the
CMASS and LOWZ samples.  Finally, the classification of $N_{\rm cp,i}$
and $N_{\rm missed,i}$ has slightly changed in DR12 relative to
DR9-DR11; see Sec.~\ref{sec:FB}.

\subsubsection{Veto Masks}
\label{sec:vetomasks}
While the basic geometry of the survey is encapsulated in the survey
mask described in the previous sections, there remain many small
regions within it where we could not have observed galaxies. Although
they are individually small, they are not randomly distributed across
the sky, and sum to a significant area, and so we exclude them from any analysis.  We
represent those regions by a set of veto masks, and remove ``randoms''
that fall within these masks. The masks are:
\begin{itemize}
\item Centerpost mask: Each Sloan plate is secured to the focal plane
  by a central bolt: no targets coinciding with the centerpost of a
  spectroscopic tile can be observed. This mask reduces the survey
  area by 0.04\%.
\item Collision priority mask: Ly$-\alpha$ quasar targets receive
  higher priority than BOSS galaxy targets in the tiling algorithm; in
  regions of only a single spectroscopic tile, BOSS galaxy targets are
  unobservable within a fibre collision radius (62$''$) of those
  targets.  Treating the high-priority quasar target locations as
  uncorrelated with the galaxy density field and neglecting any
  recovered galaxy targets in tile overlap regions, we can simply
  account for the high-priority quasars by masking a 62$''$ radius
  around each. This mask reduces the survey area by 1.5\%.
\item Bright stars mask: We mask an area around stars in the Tycho
  catalogue \citep{tycho2} with Tycho $B_T$ magnitude within [6,11.5]
  with magnitude-dependent radius
  \begin{equation}
    R = (0.0802 B_T^2 - 1.860B_T + 11.625)\,{\rm arcmin}.
  \end{equation}
  This mask reduces the area by 1.9\%.
\item Bright objects mask: The standard bright star mask occasionally
  misses some bright stars that impact the SDSS imaging data
  quality. Additionally, a small number of bright local galaxies
  saturate the imaging as well, affecting target selection in their
  outskirts. These objects were identified by visual inspection, and
  the mask radii for each object were also determined in this manner,
  ranging from 0.1\,$^\circ$ to 1.5\,$^\circ$. The number of objects
  in this mask is $\sim 125$, subtending a total area of 43.8
  deg$^2$. The list of objects is described in section~2.1 of
  \citep{Rykoff14}. This mask covers 0.4\% of the BOSS area.
\item Non-photometric conditions mask: We mask regions where the
  imaging was not photometric in $g$, $r$, or $i$ bands, the PSF
  modelling failed, the imaging reduction pipeline timed out (usually
  due to too many blended objects in a single field, caused by a high
  stellar density), or the image was identified as having any other
  critical problems. This mask reduces the area by 3.4\%.
\item Seeing cut: we discard regions where the point spread function
  full width half maximum (labeled 'PSF\_FHWM' in the catalogues) is
  greater than 2.3, 2.1, 2.0 in the $g$, $r$, and $i$ band,
  respectively.  The rationale for this cut is to decrease the
  variation of target density and properties with seeing due to the
  star galaxy separation (Eqns.~\ref{eq:sgLOWZr}, \ref{eq:sgCMASSi},
  and \ref{eq:sgCMASSz}) and $i_{\rm fib2}$ cuts.  This cut removes an
  additional 0.5\% (1.7\%) of the NGC (SGC) footprint.
\item Extinction cut: for similar reasons, we also discard areas where
  the $E(B-V)$ extinction (labeled 'EB\_MINUS\_V' in the catalogues,
  from \citealt{SFD98}) exceeds 0.15.  This cut removes an additional
  0.06\% (2.2\%) of the NGC (SGC) footprint.
\end{itemize}

In the catalogue creation pipeline, the list of targets is immediately
passed through these veto masks, so that targets in vetoed regions do
not contribute to the sector completeness calculation.  All random
galaxies within the veto regions must also be removed.  Table
\ref{tab:basic_props} shows that in total, 6.6\% (9.3\%) of the area
within the north (south) galactic cap footprint was removed by the
veto masks.

\subsection{Random Catalogue generation} \label{sec:rancat}

All of our clustering analyses make use of random catalogues with the
same angular and redshift selection functions as the data.  To produce
these catalogues, we first use the {\sc Mangle\/} ransack command to
generate one $\approx 10\times$ and two $\approx 50\times$ catalogues,
where the angular density of the random galaxies is proportional to
the completeness value in the mask for each sector\footnote{To exactly
  reproduce the officially released random catalogues, one must use the
  ransack version included in the SDSS idlutils product with version
  v5\_4\_25 or higher (Surhud More, private communication).  Random
  seeds input to ransack are provided in the catalogue generation
  scripts accompanying {\sc mksample}.}.  As the random catalogue
follows the redshift completeness per sector, it automatically
corrects for any systematic effects caused by the decrease in fiducial
exposure times starting roughly half-way through the BOSS survey. Next
we remove random galaxies using the set of veto masks described in
Sec.~\ref{sec:vetomasks}.  Only the angular coordinates of the
$10\times$ random catalogue are used to fit for angular systematic
weights; see Sec.~\ref{sec:sys}.  Since the true underlying redshift
distribution of our targets is unknown and can only be estimated from
the empirical redshift distribution, we assign redshifts to the
galaxies in the two $50\times$ random catalogues by randomly drawing
from the measured galaxy redshifts, but with a weight for each galaxy
given by $w_{\rm tot,i}$, defined in Eq.~(\ref{eq:wtot}).  This
procedure ensures that the (weighted) galaxy and random catalogues have
exactly the same redshift distribution, apart from (small)
stochasticity from the random redshift assignment. \citet{Ross12}
compare this random redshift assignment scheme with approaches that
fit a spline of varying knot number to the measured galaxy redshift
distribution, and then sample from the resulting spline directly.
Based on analysis of mock catalogues, their figure 19 demonstrates that
the former method provides the smallest bias in fits to the monopole
and quadrupole correlation function.

\section{Accounting for observational artefacts in LSS catalogues}  \label{sec:systematics}
In this section we describe in detail how we weight the targeted
galaxies when computing LSS statistics, in order to minimize the
impact of observational artefacts on our estimate of the true galaxy
overdensity field. We identify various effects that affect the
completeness of the sample, which we quantify with weights applied per
sector. These weights are a development of those presented in
\citet{And12,And14}. In particular, we discuss treatment of ``known''
redshifts from SDSS-II that were not re-observed in SDSS-III BOSS,
galaxies not observed due to fibre collisions, observed galaxies for
which a robust redshift was not obtained, and a weighting scheme to
null non-cosmological fluctuations imprinted on the catalogue by the
target selection step.  The weights described below are available for
each galaxy in the LSS catalogues.  In this section we also summarise
weights we apply to minimize our statistical error on the observed
power spectrum.

\subsection{Fibre collision corrections}
\label{sec:FB}

Galaxies that were not assigned a spectroscopic fibre due to fibre
collisions are not a random subsample of the full target sample since
they are within a fibre collision radius (62$''$) of another target.
This is potentially a large effect: in the SGC, where the coverage of
known targets from SDSS-II is lowest, approximately 20\% of galaxy
targets are in a collision group containing other CMASS or LOWZ galaxy
targets.  As a result, 5.8\% of CMASS targets and 3.3\% of LOWZ
targets were not assigned a spectroscopic fibre.

These objects preferentially occupy denser environments and therefore
have higher than average large-scale bias.  They are also more likely
than average to occupy the same dark matter halo as a neighbouring
galaxy target.  Accurate fibre collision corrections are therefore
particularly important for applications relying on the absolute value
of galaxy bias (i.e., in a comparison of the lensing and clustering
amplitude) or those that use small-scale clustering to deduce halo
occupation statistics and satellite fractions.

In the default large-scale structure catalogue that focuses on
obtaining unbiased galaxy density fields on large scales, we simply
upweight the nearest galaxy from the same target class that was
assigned a fibre to account for collided galaxies that were not
assigned fibres.  This information is tracked by incrementing a weight
$w_{\rm cp}$, labelled WEIGHT\_CP in the DR12 LSS catalogues.  The
upweighted nearest neighbour could be classified by the spectroscopic
pipeline as a good galaxy redshift, a star, or a redshift
failure. Upweighting the neighbour without reference to its
classification is the appropriate thing to do as the missed object
could be in any of these classes.

We correct 34151 (11163) CMASS targets and 4459 (4422) LOWZ targets by
nearest neighbour upweighting in the NGC (SGC).  This amounts to 5.0\%
(4.3\%) of CMASS targets in the NGC (SGC), and 1.3\% (2.9\%) of the
LOWZ targets.  The difference between the hemispheres is due both to
higher tile density in the SGC (so more fibre collisions fall in
overlap regions where they can be partially resolved) and to most of
the previously known SDSS-II redshifts falling in the NGC.

The algorithm used to generate the DR12 catalogues differs slightly from
the one used for the DR9-DR11 catalogues.  The new algorithm uses the
output from the tiling algorithm to determine membership in fibre
collision groups.  Targets with the same 'FINALN' and 'INGROUP' field
flags output from the tiling code share a collision group.  We choose
the nearest object of the same target class and collision group to
carry the weight of the unobserved target.  We also allow ``known''
galaxies to carry the weight if they are closer than all BOSS-observed
targets.  In DR9-DR11 catalogues, we did not refer to the fibre
collision group indices, but simply identified collision pairs in the
same target class if they were separated by less than 62$''$.
Nonetheless, the two algorithms select the same nearest neighbour
$\sim 94$\% of the time.

Our adopted fibre collision correction scheme neglects a few subtle cases:
\begin{itemize}
\item No corrections are applied for objects that are the only members
  of their target class in their fibre collision group, and did not
  receive a fibre.  For CMASS, this class represents 4\% of all
  targets in fibre collision groups, and 0.7\% of all CMASS targets
  overall.  Since there are more CMASS targets per unit area, this
  effect is larger for LOWZ targets: 12\% of all collided LOWZ targets
  and 1.4\% of the full sample.  Treating such collision pairs as
  unassociated is still a good approximation.  To verify this
  assumption, we examined all collision groups consisting of a single
  LOWZ target and single CMASS target, and for which we obtained both
  redshifts.  Only 11\% of such pairs had line-of-sight separations
  smaller than 50\,h$^{-1}$Mpc.
\item No corrections are applied when none of the multiple objects of
  the same target class in a fibre collision group were assigned a
  fibre.  These galaxies are treated as random incompleteness in the
  survey coverage and comprise 0.14\% of the total galaxy sample.
\item Finally, 0.3\% of targets did not receive a fibre due to
  collisions with targets other than CMASS and LOWZ but of the same
  priority.  Again we treat these missing redshifts as random.
\end{itemize}
Tables \ref{tab:fbgroups} and \ref{tab:ntilecp} provide statistics
about the distribution of CMASS and LOWZ galaxies in fibre collision
groups and how the probability of assigning a fibre to a pair of
collided galaxies in the same fibre collision group depends on the
size of the collision group.  Approximately 75\% of collided galaxies
are in a group of only two, and group sizes above four are quite rare.
In Table \ref{tab:ntilecp}, $f_{\rm fibre}$ reports the fraction of
galaxies in a collision group that received a spectroscopic fibre, as
a function of $n_{\rm tiles}$, the number of spectroscopic tiles
covering their sector.  In the remaining columns we report the
fraction of pairs of CMASS+LOWZ targets in the same collision group
for which both targets received a fibre, both globally ($c_{\rm
  pair}$), and as a function of $n_{\rm group}$.  In regions covered
by a single spectroscopic tile, only a small fraction of pairs with
$n_{\rm group} = 2$ both receive a spectroscopic fibre (4\%).  Such
pairs must be sourced from collision groups containing at least one
target of another class, oriented such that the two CMASS/LOWZ targets
in the group are separated by more than 62$''$.  As expected, for
$n_{\rm tiles} > 1$ pairs in smaller collision groups are more likely
to be resolved, and the majority of fibre collisions are removed.

Finally, to understand the impact of fibre collision corrections on
our estimate of the true galaxy density field, we examine the apparent
separation for pairs of galaxies in the same fibre collision group for
which good redshifts were obtained for both.  Fig.~\ref{fig:FBdz} shows the
distribution near $\Delta z \approx 0$, although the tails extend to
much larger separations.  We have converted redshift separations to
apparent distance separations using the fiducial cosmological model.
The observed distribution (coloured lines) can be fit by a flat
background and an exponential distribution centered on $\Delta z = 0$
(black lines).  The fraction of resolved fibre collision pairs that
are ``correlated'' (i.e., contribute to the exponential component in
the fit to the pairwise separation histogram) is 52\% for pairs of
CMASS targets and 62\% for pairs of LOWZ targets, i.e., nearly
half of fibre collision pairs are unassociated projections.
Interestingly, the width of the distribution is consistent with
$\sigma_{\rm exp} = 5.4$\,h$^{-1}$Mpc for both target classes and is
generally consistent with halo modeling expectations.

Since the choice of which galaxies are assigned fibres in a collision
group is completely random (apart from maximising the number of
targets receiving a fibre), the object not assigned a fibre is
statistically equivalent to the one we upweight, and so once
upweighted correlations at transverse separations larger than the
fibre collision scale should be unbiased.  However, correlations at
transverse separations below the collision scale will be biased, since
we are removing these small scale pairs. Additionally, these
small-scale variations will be anisotropic, and therefore likely to
have a stronger affect on the quadrupole, rather than monopole moments
of 2-point clustering statistics, for example. We therefore advocate
constructing statistics that do not apply these weights in situations
where these effects are important; see \citet{Reid14} for an example
configuration space statistic.

\begin{table}
\centering
\caption{Distribution of galaxies across fibre collision group sizes.
  The largest collisions group (not listed) contains 17 galaxy
  targets.  The first column provides the fraction of CMASS targets in
  groups with $n_{\rm group}$ CMASS targets, restricted to groups with
  at least two CMASS targets.  The second column shows the same
  calculation for LOWZ targets.  The final column lists the fibre
  collision group size distribution, where $n_{\rm group}$ includes both
  CMASS and LOWZ targets.  For consistency across the mask these
  results were computed from the LOWZ sample footprint (chunks $\geq$
  7). For reference, the fraction of galaxies that are not in any
  collision group is 77\%.}
\label{tab:fbgroups}
\begin{tabular}{llll}
\hline
$n_{\rm group}$ & $f_{\rm CMASS}$ & $f_{\rm LOWZ}$ & $f_{\rm C+L}$\\
\hline
2 & 0.7631 & 0.8456 & 0.7566\\
3 & 0.1687 & 0.1182 & 0.1726\\
4 & 0.0440 & 0.0270 & 0.0461\\
5 & 0.0146 & 0.0070 & 0.0150\\
6 & 0.0059 & 0.0010 & 0.0057\\
7 & 0.0023 & 0.0005 & 0.0022\\
8 & 0.0007 & 0.0003 & 0.0008\\
9 & 0.0003 & 0.0006 & 0.0003\\
\end{tabular}
\end{table}

\begin{table}
\centering
\caption{Fibre collision statistics for targets in regions covered by
  $n_{\rm tiles}$ spectroscopic tiles.  The second column shows the
  fraction of the total mask area covered by $n_{\rm tiles}$ tiles.
  The third column gives $f_{\rm fibre}$,  the fraction of all
  collided galaxies that were assigned a fibre.  The remaining columns
  specify the fraction of pairs of galaxy targets (CMASS + LOWZ) in
  the same collision group for which both targets received a fibre,
  both globally ($c_{\rm pair}$), and as a function of $n_{\rm
    group}$.  We use the global fraction $c_{\rm pair}$ to remove
  collided pairs and approximate the fibre collision effect in our
  mock galaxy catalogues.  We track $c_{\rm pair}$ separately for the
  NGC and SGC and for CMASS, LOWZ or combined catalogues, but in
  practice the values are similar in each case to those reported here.}
\label{tab:ntilecp}
\begin{tabular}{llllllll}
\hline
$n_{\rm tiles}$ & $f_{\rm area}$ & $f_{\rm fibre}$ & $c_{\rm pair}$ & $c_{\rm pair}(2)$ & $c_{\rm pair}(3)$ & $c_{\rm pair}(4)$ \\
\hline
1 & 0.54 & 0.561 & 0.092 & 0.042 & 0.159 & 0.142 \\
2 & 0.41 & 0.945 & 0.820 & 0.971 & 0.685 & 0.589 \\
3 & 0.05 & 0.992 & 0.966 & 0.992 & 0.985 & 0.915 \\
4 & 0.0005 & 1.000 & 1.000 & 1.000 & 1.000 & - & \\
\end{tabular}
\end{table}

\begin{figure}
\includegraphics[width=85mm]{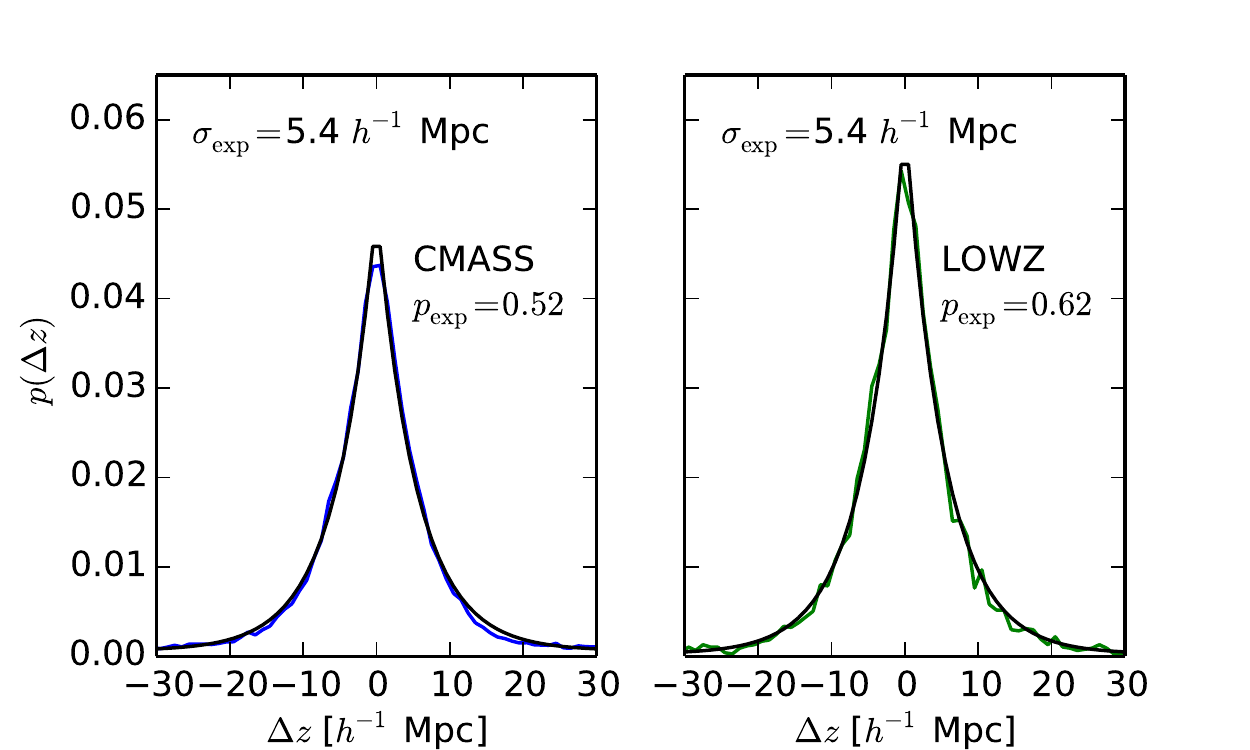} 
\caption{The probability distribution of apparent line-of-sight
  separations for pairs of galaxies in the same fibre collision group
  and for which both have good redshifts.  The left panel uses pairs
  of CMASS targets and the right panel uses pairs of LOWZ targets.
  Both distributions can be fit with the sum of a background term and
  an exponential: $a e^{-|\Delta z|/\sigma} + b$ in the range $|\Delta
  z| < 50$\,h$^{-1}$Mpc. A total of 52\% (62\%) of the CMASS (LOWZ)
  pairs contribute to the exponential term.  The best fit width
  $\sigma$ of the exponential component is 5.4\,h$^{-1}$Mpc for both
  CMASS and LOWZ targets.}
\label{fig:FBdz}
\end{figure}

\subsection{Treatment of ``known'' targets}

As the pre-observed ``known'' sample is complete (no failures are
kept), it does not match the angular distribution induced by
variations in completeness of the galaxies spectroscopically observed
by BOSS. Rather than try to model the distribution of known galaxies,
we instead subsampled these data to match BOSS completeness in each
sector, thus imposing the BOSS mask on the known galaxies. In this way
we make the sample indistinguishable from BOSS-observed targets.  In
earlier data releases (DR9-11) we also marked a fraction of the
galaxies in a 62$''$ close pair containing at least one object from
the ``known'' sample as fibre-collided; we did not apply this step in
our DR12 analysis and describe the difference in more detail in the
next section.

In DR9-DR11 catalogues we additionally marked a fraction of the galaxies
in a 62$''$ close pair containing at least one object from the ``known''
sample as fibre-collided, and assigned its weight to its nearest
neighbour.  This fraction was determined by measuring the fraction of
62$''$ BOSS targets that were fibre collision corrected in each sector.
In sectors covered only by a single spectroscopic tile all 62$''$ pairs
were collided.  The original motivation of this correction was to
impose the same fibre collision completeness on the ``known'' targets
as the BOSS targets.  In DR12 we did not apply this correction.  The
rationale was that on sufficiently large scales the nearest neighbour
upweighting scheme restores the correct clustering statistics, and so
should therefore be equivalent to using the measured redshifts.
However, we expect the effective shot noise to be larger when using
the former procedure.  Correlation function and power spectrum
analyses that marginalize over a shot noise term should be unaffected
by this choice; analyses of smaller-scale clustering should examine
this issue further.  This change is particularly important for
clustering of the LOWZ sample because of the large overlap with the
``known'' galaxy sample.

\subsection{Redshift Failures}
\label{sec:zfail}
For 1.8\% (0.5\%) of CMASS (LOWZ) targets, the spectroscopic pipeline
fails to obtain a robust redshift.  We do not necessarily expect these
to be distributed randomly with respect to e.g., plate center or
redshift, and so we again adopt a nearest neighbour upweighting scheme
to account for these objects.  Redshift failure galaxies were
permitted to be upweighted because of a nearest neighbour fibre
collision.  We therefore transfer the total weight to the nearest
neighbour of the redshift failure, incrementing a weight $w_{\rm noz}$,
labelled WEIGHT\_NOZ in the DR12 LSS catalogues.  The upweighted object
must be classified either as a good galaxy or star redshift.

In DR9-DR11 large-scale structure catalogues we removed sectors with
redshift success rates below 80\% and at least ten good redshifts; in
our DR12 catalogue we exclude troublesome observations by restricting
mask regions with PLATEQUALITY of 'good', and do not remove
the handful of sectors that would have been excluded using the DR9-11
criteria.  Upon closer examination, we found that sectors failing the
DR9-11 cut contained a small number of targets and therefore subject
to small number statistics; we checked that targets in those sectors
were drawn from plates with high redshift success rates.

In DR9-11 we searched for redshift failure neighbours to upweight only
in the same sector; in the DR12 catalogue we only consider neighbours
observed on the same plate (which spans multiple sectors) and same
date, which restricts the neighbour search to galaxies observed under
approximately the same conditions, and means the weighted number of
classified objects in each sector matches the number of targets. The
majority of close neighbours restricted to the same sector
vs. restricted to the same plate and date are the same neighbour.  The
median angular separation between galaxies without a good redshift and
their closest neighbour using the updated algorithm is 3.7$'$ (3.9$'$)
in the north (south), compared with 2.9$'$ using the sector-based
algorithm.  Total counts of redshift failures for CMASS and LOWZ
galaxies are listed in Table \ref{tab:basic_props}.

In CMASS, redshift failures are more likely to occur on faint targets
- see Fig.~\ref{fig:zdist}. In the weighting scheme described above
the neighbouring, up-weighted, galaxies are drawn from the
distribution of observed galaxies, which in turn are brighter on
average than the galaxies that failed to yield a good redshift. Given
the slight correlation of $i_{\rm fib2}$ with redshift, this
introduces a small redshift-dependent bias on the LSS catalogues. To
ameliorate this effect, we modify the redshift-failure weights such
that the {\em weighted} distribution of $i_{\rm fib2}$ of the
corrective weight matches the  $i_{\rm fib2}$ of the targets with
failed redshifts. In practice, acknowledging that an up-weighted
galaxy might be a neighbour to more than one redshift failure, we
compute $w_{noz,new} = 1 + (w_{noz,old} - 1)w_{ifib2}$, where
$w_{ifib2} = n(i_{fib2}, noz)/n(i_{fib2},cp)$  with $n(i_{fib2}, noz)$
and $n(i_{fib2},cp)$ corresponding to the green and red lines of
Fig.~\ref{fig:zdist} respectively. To avoid $w_{ifib2}$ being
dominated by Poisson noise in any given bin of $i_{fib2}$, we set
$w_{ifib2}=1$ for any bin where $n(i_{fib2}, noz)$ or $n(i_{fib2},cp)$
are less than ten. The weights are normalised such that $\sum
w_{noz,new} = \sum w_{noz,old}$. This scheme effectively transfers
weight from bright to faint neighbours of redshift-failure weights. We
only apply this extra correction to CMASS for two reasons: firstly the
LOWZ redshift-failure rate is very small ($0.5\%)$ and, secondly, we
find no significant dependence of redshift failure with $i_{\rm fib2}$
for LOWZ targets.


\subsection{Angular Systematic Weights}
\label{sec:sys}
For the DR12 data we follow the same approach as described in
\citet{Ross12} and updated in \citet{And14} to remove
non-cosmological fluctuations in CMASS target density with stellar
density and seeing.  The LOWZ targets are brighter and do not show
significant variations with these quantities, so LOWZ targets do not
require these weights.

In DR12 we update the $n_{\rm side} = 128$
HEALPix\footnote{http://healpix.jpl.nasa.gov/} stellar density map to
include all stars with $i$-band magnitudes between 17.5 and 19.9; the
map used in DR10/DR11 did not impose the 17.5 bright cut.  The two
maps also differ by a factor of the pixel area, 0.210 deg$^{-2}$.  The
functional form for $w_{\rm star}$ was also updated in DR12 to be the
inverse of a linear relation:
\begin{equation}
  w_{\rm star} (n_{\rm s}, i_{\rm fib2}) =
       (A_{i_{\rm fib2}} + B_{i_{\rm fib2}} n_{\rm s})^{-1},
  \label{eq:wstar}
\end{equation}
while in DR10/DR11 $w_{\rm star}$ was linearly dependent on $n_s$; see
\citet{Ross15} for details.  These two differences explain the changes
to the values of the $A_{i_{\rm fib2}}$ and $B_{i_{\rm fib2}}$
parameters between DR10/DR11 and DR12.  The DR12 parameter values for
$w_{\rm star}$, determined using all galaxies in the CMASS catalog
with $0.35 < z < 1.0$, are
$A_{i_{\rm fib2}} = [0.959, 0.994, 1.038, 1.087, 1.120]$ and
$B_{i_{\rm fib2}} = [0.826, 0.149, -0.782, -1.83, -2.52] \times
10^{-4}$,
computed in computed in 0.3 magnitude width $i_{\rm fib2}$ bins
centred at [20.45, 20.75, 21.05, 21.35], as in \citet{And14}.  The
parameter $w_{\rm star}$ is determined for each galaxy by first
linearly interpolating the $A_{i_{\rm fib2}}$ and $B_{i_{\rm fib2}}$
fits to derive a value at each galaxy's $i_{\rm fib2}$, and then using
Eq.~\ref{eq:wstar}. The distribution of weight values is similar in
the NGC and SGC and, overall, 93\% of CMASS galaxies have
$0.95 < w_{star} < 1.1$.

For DR10/DR11 analyses, a map of the DR8 $i$-band seeing, $S_i$, was
created by taking the mean seeing value within HEALPix pixels with
$n_{\rm side}=1024$ over the primary SDSS galaxies in the DR8
Catalogue Archive Server. For DR12, we instead directly query the
imaging data to determine the conditions estimated for each galaxy's
parent imaging field.  Per-object and per-field seeing estimates are
calculated differently. Empirically, these two methods for determining
$S_i$ differ by a factor of $\sim 0.9$.  There is also scatter between
per-field and per-object estimates of sky flux and airmass.  The DR12
galaxy and random catalogues contain fields for 'PSF\_FWHM',
'AIRMASS', 'SKYFLUX', 'EB\_MINUS\_V', and 'IMAGE\_DEPTH' if users want
to further explore systematics relationships.  In what follows, the
$i$-band seeing $S_i = {\rm PSF\_FWHM}[3]$.  For DR12 we adopted a
slightly different parameter convention from that of earlier
catalogues\footnote{Eq.~20 of \citet{And14} should state $w_{\rm see}
  (S) = 2A_{\rm see}^{-1} \left[1-{\rm erf}\left(\frac{S_i-B_{\rm
          see}}{\sigma_{\rm see}}\right)\right]^{-1}$ for the
  parameter values listed in that text.}:
\begin{equation}
  w_{\rm see} (S) = A_{\rm see}^{-1} \left[1-{\rm erf}\left(\frac{S_i-B_{\rm see}}{\sigma_{\rm see}}\right)\right]^{-1}. \label{eq:wsee}
\end{equation}
In addition, we fit the systematic relationship separately for the NGC
and SGC, again restricting the fits to objects in the CMASS LSS
catalogues with $0.35 < z < 1.0$.  The DR12 parameter values are
$A_{\rm see} =$ 0.5205 (0.5344), $B_{\rm see} =$ 2.844 (2.267), and
$\sigma_{\rm see} =$ 1.236 (0.906) for the NGC and SGC, respectively.
In DR10/DR11 we also set $w_{\rm see}(S_i > 2''.5) = w_{\rm see}(S_i =
2''.5)$; this action is no longer necessary since the DR12 veto masks
remove all area with $S_i > 2''.0$.

Finally, the application of the CMASS $z$-band star/galaxy separation
cut in the LOWZE3 sample induced a significant dependence on the
sample number density with $S_i$ that varies with the $i$-band model
magnitude; see \citet{Ross15} for details.  The systematic weight
for this sample is
\begin{eqnarray}
\ell & = & {\rm max}\left(-2, b + m (i_{\rm mod} - 16.)^{-0.5}\right) \label{eq:ell} \\
w_{\rm see,LOWZE3} & = & {\rm min}\left(5, \left(1.+(S_i-1.25)\ell \right)^{-1}\right) \label{eq:wseeL3}
\end{eqnarray}
with parameters $b = 0.875$ and $m = -2.226$, fit using all objects in
the LOWZE3 catalogue with $0.2 < z < 0.5$, including objects in chunks
$\geq 6$ in addition to the LOWZE3 targeted region, chunks 3-6.

The total angular systematic weights are simply the product of $w_{\rm
  star}$ and $w_{\rm see}$ for each object with index $i$:
\begin{equation}
\label{eq:wsystot}
w_{\rm systot,i} = w_{\rm star,i} w_{\rm see,i}
\end{equation}

\subsection{Total Galaxy Weights}
Finally, we combine the angular systematics weight $w_{\rm systot,i}$
with the fibre collision and redshift failure nearest neighbour weights
to produce a final weight for each object $i$ in the final catalog:
\begin{equation}
\label{eq:wtot}
w_{\rm tot,i} = w_{\rm systot,i} (w_{\rm cp,i} + w_{\rm noz,i} - 1).
\end{equation}
Since the default values of both $w_{\rm cp,i}$ and $w_{\rm noz,i}$
are 1, the term in parentheses conserves the total number of galaxy
targets.  This is the galaxy weighting consistent with the
construction of the LSS catalogues provided, and must be used to obtain
unbiased estimates of the galaxy density field, since this weight is
used when assigning the random galaxy redshifts; see
Sec.~\ref{sec:rancat}.

\subsection{Angular Density and Redshift Distribution}
We estimate the angular density of galaxy targets as the total number
of targets within the final LSS mask divided by the total non-vetoed
area within the sample LSS mask.  The values for each target class are
listed in the final line of Table \ref{tab:basic_props}.\footnote{Our
  calculation of the 'NBAR' field in the galaxy and random catalogues
  estimates the angular density of the sample as $A_{\rm eff}^{-1}
  \sum_i (w_{\rm cp} + w_{\rm noz} - 1)$.  Here $A_{\rm eff}$ is the
  completeness weighted area inside the mask and the sum is over all
  galaxies in the catalogue with good redshifts.  This method is
  slightly noisier since the completeness in each region is estimated
  from a finite number of galaxies.  We verified that the two methods
  agree to within 0.02\%.}  We convert this angular target density
into a three-dimensional space density through a properly normalised redshift
probability distribution:
\begin{equation}
\label{eq:pofz}
p(z_j,z_j+dz) dz \propto \sum_{z_i \in [z_j,z_j+dz]} w_{\rm tot,i} / \sum w_{\rm tot,i} ,
\end{equation}
where we sum over all objects in the catalogue with good spectroscopic
redshifts, and $w_{\rm tot,i}$ is the total weight assigned to target
$i$ to account for various observational artefacts
(Eq.~\ref{eq:wtot}).  The inclusion of $w_{\rm systot,i}$ in the estimate
for $p(z)$ accounts for any impact of the angular systematics on the
(normalised) redshift distribution, through e.g., the $i_{\rm fib2}$
dependence of the stellar weights.  However, our estimator for the
angular target density does not recover the true target density in the
absence of stars and imperfect seeing, but an average target density
over the survey footprint.  Finally, we use the fiducial cosmology to
determine the number of targets per h$^{-3}$Mpc$^3$.  The result is
shown in Fig.~\ref{fig:nofz} for all four target classes, as well as
the sum of the CMASS and LOWZ sample number densities (with duplicate
CMASS and LOWZ targets counted only once).  The CMASS+LOWZ number
density reaches a local minimum in the overlap region of
$\bar{n}(z=0.41) = 2.2 \times 10^{-4}$\,h$^{-3}$Mpc$^3$.  As reported
in the previous sections, survey incompleteness, fibre collisions,
redshift failures, and stars in the target sample all reduce the
average angular density of good galaxy redshifts compared to the
angular target density; their aggregate impact is a 10\% (4.4\%)
reduction for CMASS (LOWZ).  Finally, we compute the effective volume
$V_{\rm eff}$, which quantifies the reach of a sample for making
cosmological measurements, for the CMASS and LOWZ samples following
the same algorithm outlined in \citet{And14}, summing
over 200 redshift shells
\begin{equation}
  V_{\rm eff} = \sum\limits_{i}
  \left(\frac{\bar{n}(z_i) P_0}{1+\bar{n}(z_i) P_{0}}\right)^2 \Delta V(z_i)\,,
\end{equation}
where $\Delta V(z_i)$ is the volume of the shell at $z_i$, and we
assume that $P_0=10\,000\,h^{-3}$Mpc$^3$, which we have changed since
DR11, so the numbers are not directly comparable to \citet{And14}.  We
find $V_{\rm eff} = 5.1$\,Gpc$^3$ for CMASS and $2.3$\,Gpc$^3$ for LOWZ.

\begin{figure}
\includegraphics[width=85mm]{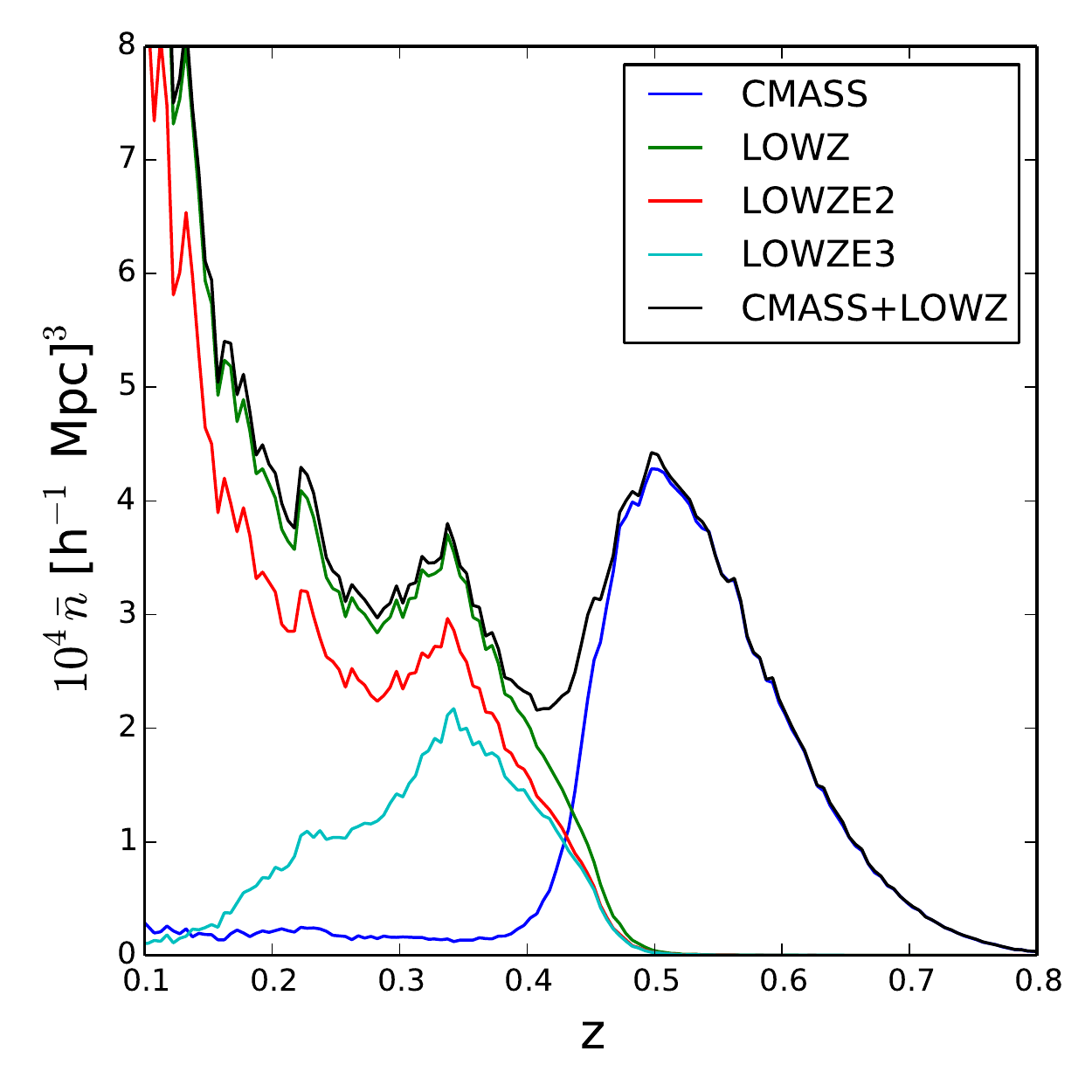} 
\caption{Number density of all four target classes assuming our
  fiducial cosmology with $\Omega_m = 0.31$, along with the sum of the
  CMASS and LOWZ number densities (black).}
\label{fig:nofz}
\end{figure}

\subsection{FKP weights}
\citet{FKP94}, hereafter FKP, showed that the optimal weighting of galaxies
as a function of redshift depends on the number density of galaxy
tracers.  The optimal weight $w_{\rm FKP}$ depends on the amplitude of
the power spectrum in the power spectrum bin of interest.  In
practice, we use the same value $P_0 = 10000$\,h$^{-3}$Mpc$^{3}$ to
estimate both the power spectrum and correlation function on all
scales.  This value of $P_0$ corresponds to the observed power
spectrum at $k \approx 0.15$\,hMpc$^{-1}$.  The field `WEIGHT\_FKP' in
the DR12 galaxy and random catalogues is given by
\begin{equation}
w_{\rm FKP,i} = \frac{1}{1+\bar{n}(z_i) P_0}
\end{equation}
for an object with redshift $z_i$, where $\bar{n}(z_i)$ is computed by
linear interpolation over bins with $\Delta z = 0.005$ starting at
$z=0$.  The $w_{\rm FKP}$ weight is optional in LSS analyses.  To utilize these
weights in a large scale structure analysis, one must weight both data
and random objects; the final weight of galaxy $i$ is therefore
$w_{\rm tot,i} w_{\rm FKP,i}$ and the final weight of random object
$j$ is $w_{\rm FKP,j}$.  If one does not use the FKP weights \citep[i.e.,
as in][]{Reid14}, consistent weightings of the galaxy and random
catalogues are $w_{\rm tot,i}$ and $w_j = 1$, respectively.

Earlier data releases adopted a different fiducial cosmology and
assumed $P_0 = 20000$\,h$^{-3}$Mpc$^3$ to compute $w_{\rm FKP}$.
\citet{PVP04} updated the analysis of \citet{FKP94} to a weighting
scheme that accounts for luminosity-dependent clustering; such weights
will be presented for the BOSS galaxy samples in a forthcoming BOSS team
paper.  However, because our target selection algorithm is so
efficient at selecting massive galaxies, the gain provided by
luminosity-dependent weights is modest for our sample.

\section{Combined catalogue creation}  \label{sec:combined}
For the purpose of providing a maximally contiguous three dimensional
density field estimate, in DR12 we provide a new catalogue that
combines the CMASS sample with the three lower redshift samples:
LOWZE2 (chunk 2), LOWZE3 (chunks 3-6), and LOWZ (chunks $\geq 7$). See
Appendix~\ref{app:lowze} for details of the LOWZE2 and LOWZE3
samples. A precise geometric description of the sky area covered by
each sample is provided in mangle mask format, constructed such that
every sector included in the CMASS mask is included in exactly one of
the LOWZE2, LOWZE3, or LOWZ footprints.  We also construct two
additional masks, one including the LOWZE2 + LOWZ sky coverage and
another including the LOWZE3 + LOWZ sky coverage.

Using those masks, we first generate a LOWZE2 catalogue including chunk
2 and chunks $\geq 7$ and a LOWZE3 catalogue including chunks $\geq 3$
using the target selection algorithms detailed in Appendix
\ref{app:lowze}.  This is possible since all the galaxies passing
LOWZE2 and LOWZE3 cuts will also pass the LOWZ cuts.  Producing a
catalogue across a larger fraction of the sky allows a more accurate
estimate of $\bar{n}(z)$ for the LOWZE2 and LOWZE3 samples (and
therefore a better means of assigning redshifts to the random galaxy
sample).  Without this step, the average density in chunk2 and chunks
3-6 would be poorly determined and could lead to erroneous
reconstruction flows towards or away from those regions in the final
combined catalogues.  As discussed in Sec.~\ref{sec:sys} and
\citet{Ross15}, there is a significant correlation between $i$-band
seeing and LOWZE3 target density which we remove using a systematic
weight given by Eq.~(\ref{eq:wseeL3}); LOWZE2 and LOWZ samples require
no systematic weight corrections.  We follow this same procedure with
some minor but important differences when combining CMASS and LOWZ
catalogues.  After full footprint data and random catalogues are produced,
we trim each catalogue back to its original targeted region (i.e.,
LOWZE2 in chunk 2, LOWZE3 in chunks 3-6, and LOWZ in chunks $\geq 7$)
using the mutually exclusive masks discussed above.
 
Our algorithm to generate the combined catalogue from the four different
samples (CMASS, LOWZ, LOWZE2, LOWZE3) is as follows:
\begin{itemize}
\item Renormalize the CMASS galaxy systematic weights $\tilde{w}_{\rm
    systot,i} \propto w_{\rm systot,i}$ such that
\begin{equation}
  1 = \frac{\sum_i \tilde{w}_{\rm systot,i}(w_{\rm cp,i} + w_{\rm noz,i} - 1)}{\sum_i (w_{\rm cp,i} + w_{\rm noz,i} - 1)}. \label{eq:wrenorm}
\end{equation}
This ensures that in the combined catalog, a CMASS target and a LOWZ
target on average have equal weight in each of the three distinct
regions.  The functional form chosen for $w_{\rm see}$ and $w_{\rm
  star}$ does not guarantee this normalisation.  Fibre collision and
redshift failure weights are left the same as in the original
CMASS-only catalogue and the parameters for the systematic weights are
identical to the ones in the CMASS-only catalogue (apart from the
renormalisation).
\item For each of LOWZ, LOWZE2, LOWZE3 samples (``LOWZX''), read in
  the targets (including those in chunks $\geq 7$), and remove objects
  already in the CMASS catalog.  Duplicate targets are 2.6\%, 2.4\%,
  and 4.4\% of the LOWZ, LOWZE2, and LOWZE3 samples, respectively.
  Fibre collision and redshift failure weights are then recomputed on
  each duplicate excluded LOWZX sample.  As in the previous catalogues,
  fibre collision and redshift failure weights are only assigned to
  other LOWZX targets (not CMASS targets).  For the LOWZE3 sample,
  systematic weights are assigned using the same parameters as the
  LOWZE3-only sample, but renormalised as in ~Eq.~\ref{eq:wrenorm}.
\item Concatenate the CMASS and LOWZX samples and compute the
  completeness of the combined sample in each sector.  The rest of the
  catalogue creation steps, i.e., random catalogue generation and $\bar{n}$
  estimation, are identical to the algorithms used for the CMASS and
  LOWZ catalogues described previously.
\end{itemize}
 
When analysing the combined catalogue, as well as allowing for any
evolution in the bias across the sample, one also has to consider the
differential bias between LOWZ and CMASS samples. Although this is
expected to be small due to the relatively benign transition from LOWZ
to CMASS \citep{Ross15}, a full exploration of this issue is left for
a forthcoming BOSS team paper.

\section{Discussion}  \label{sec:discussion}
The small statistical errors achievable on cosmological measurements
from BOSS data require removal of potential systematic issues
to an unprecedented level. Spectroscopic target selection and mask
creation are key areas where systematic problems can be introduced if
care is not taken to fully understand both. In this paper we have
presented the target selection for the three primary spectroscopic
galaxy catalogues within BOSS: LOWZ, CMASS and Sparse, and for
variations on these used for some early data. Each sample has different
sky coverage and expected redshift distribution.

We have also presented the methods used to turn the target catalogue and
redshift measurement data into galaxy and random catalogues, which
enable clustering measurements to be quickly made, as well as methods
to mitigate potential systematics. It may be that some analyses are
best done without the corrections provided - for example, it may be
cleaner for small-scale clustering analyses not to apply the
close-pair weights, but to correct in some other manner.

In addition to a number of improvements over the catalogue creation
method used for DR9, DR10 and DR11 samples we have described how we
have created a single BOSS catalog, combining CMASS and LOWZ
samples. This allows us to include some extra galaxies, and maximise
the effective volume covered by galaxies within BOSS. It also allows
us to use a binning scheme in redshift different from those of CMASS
and LOWZ, optimising our cosmological measurements.

The resulting galaxy and random catalogues, the largest in the world,
are hosted at {\tt http://www.sdss.org/dr12/} as well as supplemental
catalogue and target information. In this final release, we also provide
copies of our source code, {\sc mksample}, to reproduce the DR10,
DR11, and DR12 catalogues.  The reader should consult the source code
directly to resolve any ambiguities in our description here.

Next generation spectroscopic experiments, such as eBOSS
\citep{Daw15}, DESI \citep{levi13}, HETDEX \citep{Hill08}, 4MOST
\citep{deJong14}, WEAVE \citep{Dal14} PFS \citep{Masahiro14}, Euclid
\citep{laureijs11} and WFIRST \citep{Spergel15}, are expected to make
cosmological measurements with precision either comparable or higher
by up to an order of magnitude compared to that of BOSS, requiring a
thorough understanding and extremely careful treatment of potential
systematic effects. Although each of these future experiments have
different observing strategies, they will encounter challenges in the
process of catalogue creation similar to those of BOSS
(e.g. variations in the galaxy surface density due to galactic
extinction is an effect inherent to our observable Universe). The
lessons learned from the catalogue creation method applied within
BOSS, and described in this paper, will be of strong benefit for these
future surveys.

\section{Acknowledgements}

Funding for SDSS-III has been provided by the Alfred P. Sloan
Foundation, the Participating Institutions, the National Science
Foundation, and the U.S. Department of Energy Office of Science. The
SDSS-III web site is {\tt http://www.sdss3.org/}.

SDSS-III is managed by the Astrophysical Research Consortium for the
Participating Institutions of the SDSS-III Collaboration including the
University of Arizona, the Brazilian Participation Group, Brookhaven
National Laboratory, Carnegie Mellon University, University of
Florida, the French Participation Group, the German Participation
Group, Harvard University, the Instituto de Astrofisica de Canarias,
the Michigan State/Notre Dame/JINA Participation Group, Johns Hopkins
University, Lawrence Berkeley National Laboratory, Max Planck
Institute for Astrophysics, Max Planck Institute for Extraterrestrial
Physics, New Mexico State University, New York University, Ohio State
University, Pennsylvania State University, University of Portsmouth,
Princeton University, the Spanish Participation Group, University of
Tokyo, University of Utah, Vanderbilt University, University of
Virginia, University of Washington, and Yale University. 

We thank Eli Rykoff for providing the Bright Objects Mask described in
Section~\ref{sec:vetomasks}. 

The author order reflects the following: BR led the development of the
{\sc mksample} software package. She is followed in the author list by
two alphabetical lists of scientists who provided major contributions to
galaxy targeting and/or catalogue creation. Out of these, WJP is
corresponding author as he took-over the creation of this paper when
BR left the field of astronomy. This list is followed by an
alphabetical list of scientists who provided moderate contributions to
these topics, and who contributed to the BOSS project as a whole.

\setlength{\bibhang}{2.0em}
\setlength\labelwidth{0.0em}
\bibliography{biblio}

\appendix

\section{LOWZ Early selection algorithms}  \label{app:lowze}
As the survey progressed, these were slight changes to the targeting
pipeline.  In some instances the newer algorithm was stricter than the
one used in the past, so we simply apply the same cuts to the objects
targeted earlier as well.  One special case is the LOWZ targets in
chunks 2-6.  The star-galaxy separation algorithm for CMASS was
erroneously applied to those galaxies as well, resulting in a drastic
reduction in the target density. There are other differences, and so
we define two algorithms, LOWZE2 as that applied to chunk 2, and
LOWZE3 as that applied to chunks 3-6. In analyses thus far we have
simply eliminated these early regions from our LOWZ catalog, but we
are actively pursuing a sufficient description of that population to
robustly recover clustering measurements in those regions.  Removing
this area results in a 10\% reduction in the LOWZ survey mask
area. The distributions of the early chunks on the sky are shown in
Fig.~\ref{fig:chunks}. Chunk 2 was commissioning data, and used the
LOWZE2 version; Chunks 3-6 used LOWZE3, and chunks 7-11 used older
photometric reductions, and a different version of {\sc resolve} (see
Section~\ref{sec:imdata}). Chunk~1 was used for very early
commissioning runs and is not of sufficient uniformity to be used to
create LSS catalogues. Chunk~1 is located at DEC=0$^\circ$ in the SGC
footprint (commonly referred to as ``Stripe 82''). This area was later
reobserved with updated target selection as Chunk~11.

\begin{itemize}

\item {\bf Chunk 2}: The LOWZE2 sample had slightly different $\rcmod$
  cuts and the CMASS $i$-band star-galaxy separation cut was
  erroneously applied.  The catalogue was later trimmed to $16 <
  \rcmod$ as well. This selection yields a target density $\sim 15\%$
  lower than the nominal LOWZ target sample.
\begin{eqnarray}
\rcmod  & < & 13.4 + \cll/0.3 \\ 
|\cpp| & < & 0.2 \\
\rcmod < & 19.5 \\
\rpsf - \rcmod & > & 0.3 \\
\ipsf - \imod & > & 0.2 + 0.2(20 - \imod) \,\,.
\end{eqnarray}

\item {\bf Chunks 3-6}: The LOWZE3 sample is the same as chunk 2 but
  with a stricter $17 < \rcmod$ bound and both star-galaxy separation
  cuts. This selection yields a target density $\sim 45\%$ lower than
  the nominal LOWZ target sample.
\begin{eqnarray}
\rcmod  & < & 13.4 + \cll/0.3 \\ 
|\cpp| & < & 0.2 \\
17 & < \rcmod < & 19.5 \\
\rpsf - \rcmod & > & 0.3 \\
\ipsf - \imod & > & 0.2 + 0.2(20 - \imod) \\
\zpsf - \zmod & > & 9.125 - 0.46 \zmod \,\,.
\end{eqnarray}
\item The $\ifib < 21.5$ CMASS cut was applied to chunks 15 and above.
  Our CMASS LSS catalogue applies this cut to all chunks.

\end{itemize}

\begin{figure} \includegraphics[width=85mm]{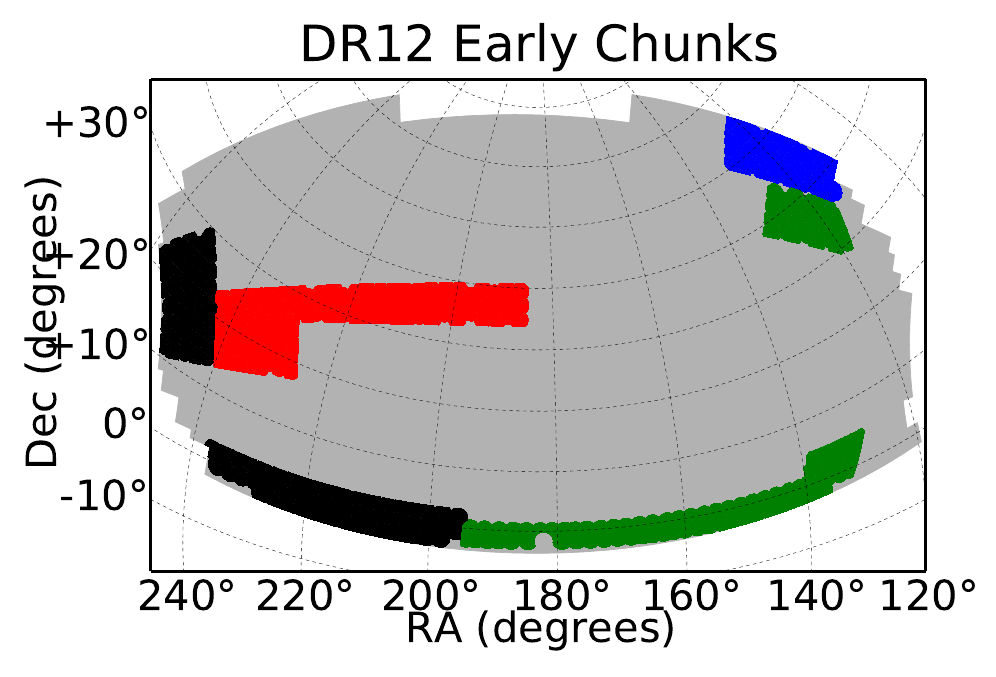} \caption{The
    location of early chunks, where targeting and/or photometric
    reduction versions differed from the later chunks.  Chunk 2
    (blue), chunks 3-4 (green), chunks 5-6 (red), and chunks 7-11
    (black) are shown. Chunks 7-11 used an early version of the
    imaging data reduction software (see Section~\ref{sec:data}).} \label{fig:chunks}
\end{figure}

\label{lastpage}
\end{document}